\documentclass[twocolumn]{aastex63}

\usepackage{makecell}
\usepackage{amsmath}
\usepackage{amsfonts}
\usepackage{amssymb}
\usepackage{bm}
\usepackage{graphicx}
\usepackage{xcolor}

\newcommand{\be}{\begin{equation}}
\newcommand{\ee}{\end{equation}}

\newcommand{\safeincludegraphics}[2][]{%
  \IfFileExists{#2}{\includegraphics[#1]{#2}}{%
    \fbox{\parbox{0.85\linewidth}{\centering Missing figure file: \texttt{\detokenize{#2}}}}%
  }%
}

\begin{document}

\title{GRMHD Simulations of Accreting Proto-Magnetars I.\\ Implications for Gamma-Ray Burst Jets and Energetic Explosions}

\author{Tejas Prasanna}
\affiliation{Department of Physics and Columbia Astrophysics Laboratory, Columbia University, New York, NY 10027, USA}
\author{Kyle Parfrey}
\affiliation{Princeton Plasma Physics Laboratory, Princeton, NJ 08540, USA}
\author{Brian D. Metzger}
\affiliation{Department of Physics and Columbia Astrophysics Laboratory, Columbia University, New York, NY 10027, USA}
\affil{Center for Computational Astrophysics, Flatiron Institute, 162 5th Ave, New York, NY 10010, USA}
\author{Ore Gottlieb}
\affil{Department of Physics and Kavli Institute for Astrophysics and Space Research,
Massachusetts Institute of Technology, Cambridge, MA 02139, USA}
\author{Andrei Beloborodov}
\affiliation{Department of Physics and Columbia Astrophysics Laboratory, Columbia University, New York, NY 10027, USA}
\author{Koushik Chatterjee}
\affiliation{Canadian Institute for Theoretical Astrophysics, University of Toronto, 60 St. George Street, Toronto, ON M5S 3H8, Canada}

\begin{abstract}
Newly formed, rapidly rotating, strongly magnetized neutron stars
(``millisecond proto-magnetars'') are promising central engines for
gamma-ray bursts (GRBs) and luminous supernovae. Although often modeled
in isolation, they can be born surrounded by accretion disks in stellar
collapse, neutron-star mergers, or accretion-induced collapse. We
present axisymmetric GRMHD simulations of hyperaccretion onto such
objects, including a physical equation of state and charged-current
weak interactions. Holding the weakly magnetized accretion torus fixed,
we vary the stellar dipole field strength to span
crushed-magnetosphere, magnetically channeled accretion, and
centrifugal-propeller regimes, and compare the results with an otherwise
similar accreting black hole.

Accretion compresses the stellar magnetosphere and opens additional
magnetic flux, producing relativistic jet powers that exceed
isolated-dipole spin-down estimates by factors of a few to
$\sim 10$. Even while the magnetosphere remains compressed against the
stellar surface, stronger fields increasingly impede accretion and
enhance the outflows. Channeled-accretion models show strong jet
variability driven by plasmoid eruptions and intermittent
magnetospheric accretion, whereas the propeller model produces a
steadier, more powerful jet and a rapid spin-down torque. The
disk--magnetosphere interaction also regulates how efficiently the
neutron star grows in mass and whether it spins up or spins down; near
spin equilibrium, inefficient accretion can delay collapse to a black
hole relative to estimates based on the external mass-supply rate.
Accreting proto-magnetars can therefore power relativistic jets and
baryon-rich outflows with energetics comparable to those inferred for
long GRBs and GRB-supernovae, respectively. A companion paper explores
the implications of the disk--magnetosphere interaction for
neutron-rich ejecta and $r$-process nucleosynthesis.
\end{abstract}

\section{Introduction}

Young, rapidly rotating neutron stars (NS) with strong large-scale magnetic fields $\gtrsim 10^{15}$ G are promising central engines of gamma-ray bursts (GRBs), luminous supernovae, and electromagnetic counterparts to compact object mergers
\citep[e.g.,][]{Usov92,Blackman&Yi98,Kluzniak&Ruderman98,Thompson+04,Metzger+11a,Kasen&Bildsten10,Woosley10,Beniamini+17}.  Such ``millisecond proto-magnetars'' can form during the core-collapse of rotating massive stars (e.g., \citealt{Mosta+14,Obergaulinger&Aloy21}), the accretion-induced collapse (AIC) of rapidly spinning white dwarfs (e.g., \citealt{Dessart+06,Metzger+08a}), or as leftover remnants following the merger of binary NSs (e.g., \citealt{Bucciantini+12,Metzger+18,Kiuchi+24}).  For spin periods $P_{\star} \sim 1~{\rm ms}$, the magnetar possesses a large rotational energy $E_{\rm rot} \gtrsim 10^{52}~{\rm erg}$, which can be extracted through magnetized outflows on timescales as short as tens of seconds.  Although the origin of such strong, large-scale magnetic fields remains
uncertain, dynamo action powered by differential rotation provides a
plausible mechanism in the fastest-spinning neutron stars
\citep{Duncan&Thompson92,Raynaud+20,Most&Quataert23,
Kiuchi+24,combi_aic_2025}.

In magnetar models for GRBs, a relativistic jet\footnote{Throughout this paper we use the term ``jet'' to denote a relativistic, magnetically powered polar outflow launched by the proto-magnetar, even when its opening angle near the engine is large.  This terminology is motivated by the expected role of this component as the GRB-powering outflow: in a collapsar, for example, the stellar envelope can help collimate an initially wide-angle relativistic wind
into a narrower jet on larger scales \citep[e.g.,][]{Bromberg+14,Gottlieb+22}.  We reserve the term ``wind'' for the slower, baryon-rich disk or propeller outflow, unless otherwise specified.} is powered by the electromagnetic luminosity carried along open polar magnetic field lines which are relatively free of baryons and hence achieve high magnetization $\sigma \gg 1$ and ultra-relativistic speeds \citep{Bucciantini+06,Komissarov&Barkov07,Uzdensky&MacFadyen07,Metzger+11a,Combi&Siegel23,combi_aic_2025,Desai+26}.  More heavily baryon-loaded outflows emerge at lower latitudes, where
neutrino heating and centrifugal acceleration enhance the mass-loss
rate and kinetic power
\citep{Thompson+04,Metzger+18b,Ciolfi+20,
Prasanna+23,Curtis+24,Prasanna+25}.  On larger scales, the magnetar jet and wind can interact with and boost the energy of the supernova or merger ejecta (e.g., \citealt{Thompson+04,Metzger&Piro14,Kasen+16,Chen+16,Suzuki&Maeda21}). However, most theoretical proto-magnetar models neglect continued
accretion and instead assume that the star spins down in isolation.

In realistic formation channels, however, continued accretion onto the
newly formed neutron star is likely to be common. In collapsars,
high-angular-momentum stellar material can circularize into a disk on
timescales of seconds \citep{MacFadyen&Woosley99,Zhang+08,
Burrows+23,Gottlieb+24}. Neutron-star mergers produce massive,
neutrino-cooled tori, while rapidly rotating white-dwarf AIC can likewise
leave a compact disk around the remnant
\citep{Lee&RamirezRuiz07,Fernandez&Metzger13,Fahlman&Fernandez18,
Dessart+06,Metzger+08a,Cheong+25,combi_aic_2025}. Typical disk masses $M_{\rm d} \sim 10^{-2}-1\,M_\odot$ and accretion
timescales $\sim 0.1$--$10~{\rm s}$ imply hyperaccretion rates up to $\dot{M} \sim M_{\odot}$ s$^{-1}$ during the first seconds after the collapse or merger. As a result, one must in general consider the coupled interaction between the magnetized rotating star and a powerful accretion flow.

Previous analytic and numerical studies show that accretion can
qualitatively alter both the power and time dependence of a
proto-magnetar engine \citep{Piro&Ott11,Gompertz+14,Metzger+18b}.
By pushing the magnetosphere inside the light cylinder, the disk can
open additional stellar magnetic flux and enhance the spin-down
luminosity relative to that of an isolated dipole
\citep{Parfrey+16,Metzger+18b}. The resulting increase in early energy
injection can affect the dynamics and nucleosynthesis of the surrounding
supernova ejecta, including the production of radioactive
$^{56}{\rm Ni}$ \citep{Nakamura+01,Suwa&Tominaga15}.
Accretion can also make the engine strongly time dependent:
intermittent penetration of the magnetosphere may modulate the jet
power and baryon loading, while the competing accretion and
electromagnetic torques govern the longer-term spin evolution
\citep{Bernardini+13,Gompertz+14,Metzger+18b}.
These changes at the engine can subsequently shape the high-energy
radiation produced by dissipation at much larger radii
\citep{Giannios&Spruit07,Beloborodov10}.

Here, we present axisymmetric GRMHD simulations of neutrino-cooled
accretion onto rapidly spinning proto-magnetars. Holding the initial
torus and its outer mass-supply rate approximately fixed, we vary the
stellar dipole strength to follow the progression from a crushed
magnetosphere, through magnetically channeled accretion and approximate
spin equilibrium, to the centrifugal propeller regime. We compare this
sequence directly with an otherwise similar black-hole accretion flow.

This paper, hereafter Paper~I, focuses on how the
disk--magnetosphere interaction regulates the power, magnetization,
variability, and angular-momentum flux of the relativistic polar jet,
as well as the accretion and spin evolution of the neutron star. The
same interaction also regulates the mass, velocity, entropy, and
electron fraction of the baryon-rich ejecta. Those consequences and
their implications for $r$-process nucleosynthesis are explored in the
companion Paper~II.

In Sec.~\ref{sec:analytic} we review the analytic estimates that define
the relevant physical scales. In Sec.~\ref{sec:methods} we describe the
numerical methods and model suite. In Sec.~\ref{sec:results} we present
the disk--magnetosphere interaction, jet energetics and variability,
and neutron-star torques. In Sec.~\ref{sec:applications} we discuss
applications to collapsars, neutron-star mergers, and AIC. We summarize
our conclusions and limitations in Sec.~\ref{sec:conclusions}.

\section{Accreting Proto-magnetars}
\label{sec:analytic}

\begin{figure}[t]
    \centering
    \safeincludegraphics[width=0.5\textwidth]{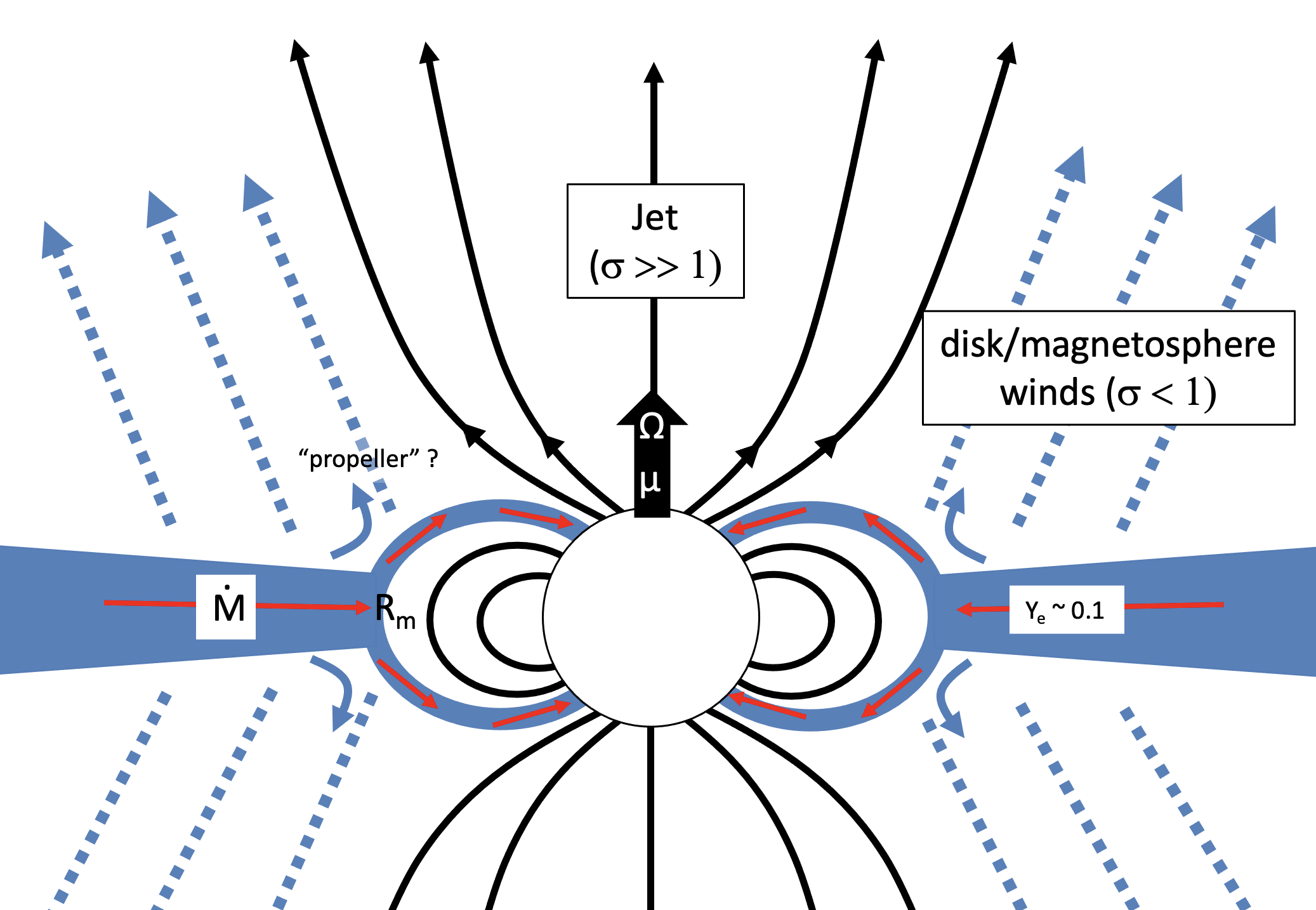}
    \caption{
    Schematic illustration of an accreting proto-magnetar central engine.
    An NS with rotational angular frequency $\Omega_{\star}$ (spin period $P_{\star} = 2\pi/\Omega_{\star})$ and magnetic dipole moment $\mu_{\star}$
    accretes from a neutron-rich disk at a high rate $\dot{M} \gtrsim 10^{-2}\,M_{\odot}$ s$^{-1}$.
    The accretion flow is truncated where it encounters the stellar magnetosphere,
    at approximately the Alfv\'en radius $R_{\rm m}$.
    Open polar magnetic field lines power a relativistic jet, whose power is enhanced relative to that of isolated dipole spin-down by the additional poloidal magnetic flux opened by the interaction between the inner disk and the rotating magnetosphere.  The star--disk interaction can also enhance baryon-rich disk/propeller winds, whose composition and nucleosynthetic implications are the subject of Paper~II.
    }
    \label{fig:cartoon}
\end{figure}

We consider an accretion disk of mass $M_{\rm d} \sim 10^{-2}-1\,M_{\odot}$ formed around a newly born compact object of mass $M_{\star}$ in stellar core-collapse, AIC, or a NS merger event, with a characteristic radius $R_d\gtrsim30\,r_g$, where $r_{\rm g} \equiv GM_{\star}/c^{2}$ is the gravitational radius (see Fig.~\ref{fig:cartoon} for a schematic illustration).  The accretion rate can be estimated as 
\begin{eqnarray}
\dot{M}\sim \frac{M_{\rm d}}{t_{\rm visc}} \approx 0.05 \ M_\odot\,{\rm s}^{-1} \left(\frac{M_{\rm d}}{0.2\,M_\odot}\right) \times \nonumber \\
\left(\frac{R_{\rm d}}{50\,r_{\rm g}}\right)^{-3/2}\left(\frac{M_{\star}}{2\,M_\odot}\right)^{-1}
\left(\frac{h/r}{0.3}\right)^2
\left(\frac{\alpha}{0.01}\right),
\label{eq:Mdot}
\end{eqnarray}
where $t_{\rm visc}$ is the viscous timescale within an $\alpha$ viscosity prescription \citep{Shakura1973}, and $h/r$ is the disk aspect ratio scaled to a characteristic value for thick neutrino-cooled disks (e.g., \citealt{Narayan+01}).

At such high accretion rates, the inner regions of the disk can become dense and hot enough that weak interactions---primarily electron
and positron captures on free nucleons---proceed faster than the local
accretion time (e.g., \citealt{Beloborodov03}). In this regime, the disk cools efficiently and its midplane electron fraction $Y_{\rm e}$ is driven toward a low equilibrium value through a self-regulated balance between weak interactions, neutrino cooling, and electron degeneracy \citep{Chen&Beloborodov07,Siegel&Metzger17,HernandezMorales&Siegel25}. This neutron-rich transition occurs interior to an ignition radius
$R_{\rm ign}$, defined by the condition that weak interactions become
faster than the local inflow time. Equivalently, at a specified disk
radius $r \lesssim R_{\rm d}$, the transition occurs above a critical ``ignition''
accretion rate, which can be expressed as \citep{Metzger+08}
\begin{eqnarray}
\dot{M}_{\rm ign} &\approx& 5\times 10^{-3}\ M_\odot\,{\rm s}^{-1} \nonumber \\
&&\times \left(\frac{r}{50\,r_{\rm g}}\right)^{5/6}
\left(\frac{M_{\star}}{2\,M_\odot}\right)^{4/3}
\left(\frac{\alpha}{0.01}\right)^{5/3}.
\label{eq:Mdotign}
\end{eqnarray}
For $\dot{M} \gg \dot{M}_{\rm ign}$, the disk midplane is dense and moderately electron-degenerate, suppressing positron captures and
driving the electron fraction towards characteristic values
$Y_{\rm e} \approx 0.1-0.2$, largely independent of the initial composition
(e.g., \citealt{Chen&Beloborodov07,Siegel&Metzger17}). 

For a magnetized NS, the accretion disk is truncated roughly where
magnetic stresses balance the stresses of the inflowing disk. A
commonly used order-of-magnitude estimate is the spherical Alfv\'en
radius \citep{Davidson1973,Ghosh&Lamb79},
\begin{eqnarray}
R_{\rm m} \approx \left(\frac{\mu_{\star}^4}{2G M_{\star} \dot{M}^2}\right)^{1/7} \approx 
4.5\,r_{\rm g}\,\left(\frac{B_{\star}}{2\times 10^{15}\ {\rm G}}\right)^{4/7} \nonumber \\
\times\left(\frac{R_\star}{12\ {\rm km}}\right)^{12/7} 
\left(\frac{M_{\star}}{2\,M_\odot}\right)^{-8/7}
\left(\frac{\dot{M}}{0.1\,M_\odot\,{\rm s}^{-1}}\right)^{-2/7},
\label{eq:Rm}
\end{eqnarray}
where $\mu_{\star} = B_{\star}R_{\star}^{3}$ is the magnetic dipole moment of the magnetar of radius $R_{\star}$ and equatorial surface field strength $B_{\star}$.  The normalization and scaling can differ for a geometrically thick,
disk-confined magnetosphere; indeed, our simulations yield a somewhat
shallower dependence on $B_\star$ (Fig.~\ref{fig:MagRadScaling}).

Absent accretion, the magnetar will spin down via a magnetically dominated wind.  For an aligned rotator in the force-free limit\footnote{Early in the magnetar evolution, the wind is not force-free as a result of neutrino-driven mass-loss along the open magnetic field lines.  These inertial effects open additional magnetic field lines and can substantially increase the wind power relative to the force-free spin-down rate (e.g., \citealt{Thompson+04,Metzger+07,Prasanna+22,Prasanna+23,Desai+26}), even absent an accretion disk.}, the spin-down luminosity can be written as \citep{Gruzinov05,Spitkovsky06}
\begin{eqnarray}
&\dot{E}_{0}&
\simeq
k\frac{\mu_{\star}^2\Omega_{\star}^4}{c^3} \approx 4.3\times 10^{49}\ {\rm erg\ s^{-1}} \times
\nonumber \\
&&\left(\frac{k}{0.25}\right)\left(\frac{B_{\star}}{10^{15}\ {\rm G}}\right)^2 \left(\frac{R_\star}{12\ {\rm km}}\right)^6
\left(\frac{P_{\star}}{1\ {\rm ms}}\right)^{-4}
\label{eq:Edotsd}
\end{eqnarray}
where $\Omega_{\star}$ and $P_{\star} = 2\pi/\Omega_{\star}$ are the rotational frequency and spin-period of the NS, respectively. Here $k$ accounts for the mapping between the reported fluid-frame
surface field and the magnetic dipole moment entering the asymptotic
spin-down solution. For our stellar compactness, a fixed dipole moment
corresponds to a fluid-frame surface field approximately twice the
flat-space value. Because $\dot E_0\propto\mu_\star^2$, expressing the
spin-down luminosity in terms of the reported fluid-frame field gives
$k\simeq0.25$.

Accretion can increase the spin-down luminosity of the magnetar jet by increasing the fraction of open magnetic flux \citep{Parfrey+16,Metzger+18b}. This occurs if the magnetospheric radius $R_{\rm m}$ (Eq.~\eqref{eq:Rm}) is located inside the light-cylinder radius,
\begin{eqnarray}
R_{\rm LC} =
\frac{c P_{\star}}{2\pi}
\simeq
16 \,r_{\rm g}
\left(\frac{P_{\star}}{1\,{\rm ms}}\right)
\left(\frac{M_{\star}}{2\,M_\odot}\right)^{-1}.
\label{eq:RL}
\end{eqnarray}
When $R_{\rm m} < R_{\rm LC}$, the disk--magnetosphere interaction enhances the
spin-down power relative to the isolated dipole rate by a factor
$\simeq (R_{\rm LC}/R_{\rm m})^2$.  Combining Eqs.~\eqref{eq:Rm}--\eqref{eq:RL} gives an estimate of the enhanced electromagnetic
spin-down power available to the relativistic jet,
\begin{eqnarray}
\dot{E} &\simeq& \dot{E}_0 \left(\frac{R_{\rm LC}}{R_{\rm m}}\right)^2 \approx 2 \times 10^{51}\ {\rm erg\ s^{-1}}\,\left(\frac{B_\star}{2 \times 10^{15}\ {\rm G}}\right)^{6/7} \nonumber \\
&\times& \left(\frac{R_\star}{12\ {\rm km}}\right)^{18/7}
\left(\frac{P_\star}{1\ {\rm ms}}\right)^{-2} 
\nonumber \\
&\times&\left(\frac{M_\star}{2M_\odot}\right)^{2/7}
\left(\frac{\dot{M}}{0.1\,M_\odot\ {\rm s}^{-1}}\right)^{4/7},
\label{eq:Edot}
\end{eqnarray}
where the numerical normalization assumes $k=0.25$; this expression
is valid for $R_\star<R_{\rm m}<R_{\rm LC}$.  For high enough accretion rates, such that $R_{\rm m} \propto \dot{M}^{-2/7}$ approaches the NS surface ($R_{\rm m} \rightarrow R_{\star})$, the spin-down power approaches that of a split-monopole.  
However, in this regime the polar outflow may become heavily mass-loaded or confined by the
accretion flow, reducing the fraction of the spin-down power that
emerges in a high-$\sigma$ relativistic jet.

The efficiency with which disk material is able to penetrate the magnetosphere depends on the rotation rate of the magnetar.  One can define the corotation radius, 
\begin{eqnarray}
R_{\rm c}
=
\left(\frac{G M_{\star}}{\Omega_{\star}^2}\right)^{1/3}
\simeq
6.3 r_{\rm g}
\left(\frac{P_{\star}}{1\,{\rm ms}}\right)^{2/3}
\left(\frac{M_{\star}}{2\,M_\odot}\right)^{-2/3},
\label{eq:Rcor}
\end{eqnarray}
as the location in the disk where the Keplerian orbital frequency $\Omega = (GM_{\star}/r^{3})^{1/2}$ matches that of the NS rotation, $\Omega_{\star}$.  Schematically, if $R_{\rm m}\lesssim R_{\rm c}$, accretion can proceed onto the star along closed magnetic field lines, whereas for a magnetosphere maintained outside corotation,
$R_{\rm m}\gtrsim R_{\rm c}$, the system enters the propeller regime 
\citep{Illarionov&Sunyaev75,Romanova2004}, leading to suppressed and/or intermittent accretion onto the NS surface and more powerful outflows than in the accretion regime.

\begin{figure}
\centering
\safeincludegraphics[width=\columnwidth]{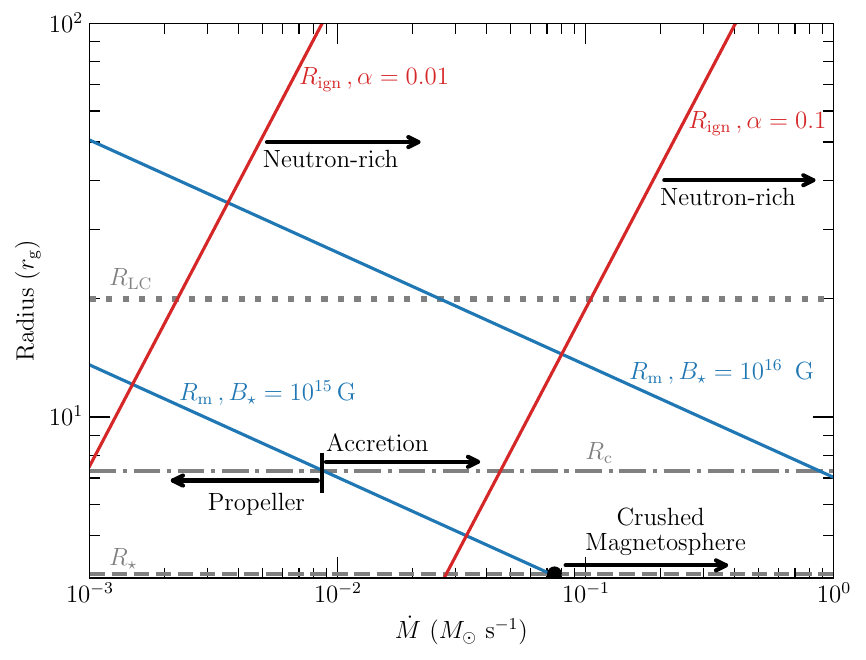}
\caption{
Critical radii for accretion onto a millisecond proto-magnetar as a function of the mass accretion rate $\dot{M}$, shown in units of $r_{\rm g}=GM/c^2$ for $M_{\star}=2M_\odot$, $P_{\star} =1.26\,{\rm ms}$, and $R_\star=12\,{\rm km}$.
Solid blue lines show the magnetospheric radius $R_{\rm m}$ from Eq.~(\ref{eq:Rm}) for surface dipole fields $B_{\star}=10^{15}\,{\rm G}$ and $10^{16}\,{\rm G}$.
Solid red lines show the ignition radius $R_{\rm ign}$ obtained by inverting Eq.~(\ref{eq:Mdotign}) for $\alpha=0.01$ and $0.1$.
Horizontal dashed, dash-dotted, and dotted lines mark the neutron-star radius $R_\star$, corotation radius $R_{\rm c}$, and light-cylinder radius $R_{\rm LC}$, respectively.
The region to the lower right of the ignition boundary is neutron-rich, $\dot{M}\gtrsim\dot{M}_{\rm ign}$.
For $R_{\rm m}\lesssim R_{\rm LC}$, disk accretion enhances the open magnetic flux and spin-down power relative to the usual isolated dipole rate (Eq.~\eqref{eq:Edotsd}), while the transition across $R_{\rm m}\simeq R_{\rm c}$ separates the propeller regime, $R_{\rm m}\gtrsim R_{\rm c}$, from accretion onto the surface, $R_{\rm m}\lesssim R_{\rm c}$.
}
\label{fig:critical_radii}
\end{figure}

As summarized in Fig.~\ref{fig:critical_radii}, the same region of
parameter space that produces neutron-rich disks,
$\dot{M}\gtrsim \dot{M}_{\rm ign}$, also spans the key
disk--magnetosphere regimes for millisecond proto-magnetars.  At fixed
$\dot{M}$ and $P_\star\sim1\,{\rm ms}$, increasing the stellar dipole
field moves the system from a crushed-magnetosphere state,
$R_{\rm m}\lesssim R_\star$, to magnetically channeled accretion,
$R_\star \lesssim R_{\rm m}\lesssim R_{\rm c}$, and finally to the
centrifugal propeller regime, $R_{\rm m}\gtrsim R_{\rm c}$.  Throughout
much of this sequence $R_{\rm m}<R_{\rm LC}$, so disk compression of the
stellar magnetosphere can open additional magnetic flux and enhance the
electromagnetic luminosity to
$\dot E\sim10^{50}$--$10^{52}\,{\rm erg\,s^{-1}}$
(Eq.~\ref{eq:Edot}), comparable to that required of GRB central
engines.  Thus, for the accretion rates relevant to neutron-rich disks
and the magnetic fields capable of powering relativistic jets, the
inner flow can naturally transition between crushed, accreting, and
propeller states.  Fully capturing this
disk--magnetosphere--jet interaction requires global GRMHD simulations,
which we now describe.

\section{Numerical Methods}
\label{sec:methods}
We perform GRMHD simulations of accreting neutron stars (NSs) and
black holes (BHs) using the H-AMR code \citep{Liska2022}. We employ
the extension developed for NS magnetospheres \citep{Chatterjee2026},
which self-consistently evolves MHD and magnetically dominated
force-free regions within a single calculation
\citep{Parfrey&Tchekhovskoy+17,Parfrey&Tchekhovskoy24}.
We further modify the code to incorporate a general tabular equation
of state (EOS) and passive-scalar evolution of the electron fraction
\citep{Berthier2021}, as described below.

The ideal GRMHD equations include baryon conservation,
\begin{equation}
    \nabla_{\mu} \left( \rho u^{\mu} \right)=0,
\end{equation}
energy-momentum conservation, 
\begin{equation}
\label{energy_eqn}
    \nabla_{\mu} T^{\mu}_{ \ \  \nu}= Qu_{\nu},
\end{equation}
and Maxwell's equations,
\begin{equation}
    \nabla_{\nu} \,  ^{*}F^{\mu \nu}=0,
\end{equation}
where 
\begin{equation}
    T^{\mu \nu}= \left(\rho + \frac{\epsilon + p + b^2}{c^2} \right)u^{\mu}u^{\nu} + \left( p + \frac{b^2}{2} \right)g^{\mu \nu} - b^{\mu}b^{\nu}
\end{equation}
is the total energy-momentum tensor.  Here, $\rho$, $\epsilon$, and $p$ are the fluid-frame mass density, internal energy density, and gas pressure respectively; $u^{\mu}$ is the fluid four-velocity, $g^{\mu \nu}$ is the metric tensor, $b^{\mu}$ is the fluid-frame magnetic four-vector in Lorentz-Heaviside units, and $^{*}F^{\mu \nu}=b^{\mu}u^{\nu}-b^{\nu}u^{\mu}$ is the dual of the electromagnetic field tensor. The source term $Q(\rho,T,Y_{\rm e})$ in Eq. \eqref{energy_eqn} is the net heating/cooling rate per unit volume in the rest frame of the fluid due to neutrino emission, as described in Sec.~\ref{sec:weak}. Weak interactions also change the electron fraction of the fluid, which is evolved as a separate passive scalar:
\begin{equation}
    \label{YE_evol_eqn}
     \nabla_{\mu} \left(Y_{\rm e} \rho u^{\mu} \right)=\rho R,
\end{equation}
where $R(\rho,T,Y_{\rm e})$ is the specific lepton emission/absorption rate in the fluid rest frame (Sec.~\ref{sec:weak}).

We also evolve a separate passive scalar $\mathcal{F}$ obeying,
\begin{equation}
    \nabla_{\mu} \left(\mathcal{F} \rho u^{\mu} \right)=0,
\end{equation}
which we use to separate disk material from the background atmosphere in the neutron star simulations \citep{Parfrey&Tchekhovskoy+17}.

We adopt a fixed Kerr metric with spin parameter $a=1/3$ for both the
BH and NS simulations, with the NS surface imposed at
$R_\star=4\,r_{\rm g}$. Thus, the NS calculation uses a Kerr exterior
rather than a self-consistent rotating stellar spacetime. This
approximation permits a controlled comparison between compact objects
with the same mass and angular momentum while isolating the effects of
the stellar surface and anchored magnetic field. A constant central mass is a good approximation because for accretion rates of interest, the NS or BH would grow by $\lesssim 1\%$ over the duration of the simulations.  Likewise, our chosen torus mass ($0.2M_{\odot}$; Sec.~\ref{sec:torus}) is sufficiently small that self-gravity is not expected to play an important role in the evolution of the system.

\subsection{Computational Grid and Boundary Conditions}
We employ a 2D spherical polar grid with $N_r$ cells in the radial direction, and $N_{\theta}$ cells in the polar direction. The radial domain extends from the inner boundary at a radius of
$R_{\rm in}=R_\star=4\,r_{\rm g}$ in the NS models and from
$R_{\rm in}= 1.8\,r_{\rm g}$ in the BH models to
$R_{\rm out}=10^4\,r_{\rm g}$. The radial grid is logarithmically spaced, and the angular grid spanning $[0,\pi]$ is uniformly spaced.  All our simulations adopt $N_r = 512$, $N_{\theta} = 256$. In Sec.~\ref{sec:torus}, we show that this choice allows a sufficient resolution of the fastest-growing MRI mode.   

We use reflective boundary conditions in the $\theta$ direction and
apply the same outflow prescription for the hydrodynamic primitive
variables at the inner and outer radial boundaries of the NS and BH
models, copying their values from the nearest active cell into the
ghost cells. At the inner boundary, this permits nearly unimpeded
inflow into the compact object. For the NS models, this treatment is
physically motivated by the efficient neutrino cooling of the
accretion layer at the rates of interest,
$\dot{M}\gtrsim10^{-3}\,M_\odot\,{\rm s}^{-1}$
\citep{Chevalier93,Combi+25}. 
In the NS models we apply electromagnetic inner boundary conditions that anchor the stellar magnetic field into the surface. The magnetic field normal to the surface is held fixed while its tangential components evolve freely. The tangential electric fields (EMFs) that constitute the fluxes for the induction equation are found from the spacetime-appropriate form of $\boldsymbol{E} = - \boldsymbol{v}\times\boldsymbol{B}$, where here $v^r = v^\theta = 0$, $v^\phi = \Omega_\star$.  
%

We impose density floors in low-density, strongly magnetized regions
to limit the maximum magnetization and maintain numerical stability.
For the NS simulations, we enforce a maximum $b^2/\rho=50$ at the
stellar surface and a maximum $b^2/\rho=25$ outside the light cylinder,
with a smooth interpolation between these limits. For the BH
simulations, we impose a maximum $b^2/\rho=25$ throughout the domain.
Here the ceiling is imposed on $b^2/\rho$, whereas elsewhere we
define the relativistic magnetization as
$\sigma=b^2/(h\rho)$ where $h$ is the specific enthalpy.  This enables the relativistic ``jet'' to be distinguished from more
weakly magnetized ($\sigma < 1$) accreting or unbound disk material
(see also Fig.~\ref{fig:cartoon}). We have verified that, over the
radii analyzed in this paper, the total electromagnetic power carried
by material with $\sigma>1$ is insensitive to the adopted density-floor
prescription. The floors do, however, affect how this power is distributed at higher
magnetization, particularly for $\sigma\gtrsim5$. We therefore interpret
the simulations as measuring the electromagnetic power supplied to the
relativistic funnel, rather than the absolute high-$\sigma$ distribution
or detailed terminal Lorentz-factor structure of the jet.

In a physical proto-magnetar, baryon loading is regulated by neutrino
heating in the stellar atmosphere and depends on the neutrino luminosity
and cooling age \citep{Thompson+04,Metzger+11a,Desai+26}. Our adopted
floors permit peak values $\sigma\simeq20$ in both the accreting jets
and the disk-free model \texttt{NS-B1e15-nodisk}, within the broad range
$\sigma\sim10$--$10^3$ expected several seconds after NS birth. For the
accreting magnetars of interest, irradiation of the NS surface by
neutrinos from the hot accretion flow may further regulate the baryon
loading of the jet and limit its maximum magnetization
(\citealt{Metzger+18b}; see Sec.~\ref{sec:applications} for further
discussion).

\subsection{Equation of State}

We employ the general Helmholtz EOS \citep{Timmes2000}, tabulated as a
function of density $\rho$, temperature $T$, and electron fraction
$Y_{\rm e}$. We assume nuclear statistical equilibrium among free
neutrons (n), free protons (p), and $\alpha$ particles, whose relative abundances we determine as a function of $(\rho,T,Y_{\rm e})$ by numerically solving the Saha equation simultaneously with baryon number conservation and charge conservation as follows \citep{Siegel2018}:
\begin{align}
    n_{\rm p}^2 n_{\rm n}^2 &= 2n_{\alpha} \left(\frac{m_{\rm b} k_{\rm B} T}{2\pi\hbar^2}\right)^{9/2} \exp \left( -Q_{\alpha}/k_{\rm B}T\right) \\
    n_{\rm b} &= n_{\rm n} + n_{\rm p} + 4n_{\alpha} \\
    n_{\rm b} Y_{\rm e} &= n_{\rm p} + 2n_{\alpha},
\end{align}
where $n_{\rm b}$, $n_{\rm n}$, $n_{\rm p}$, and $n_{\alpha}$ are number densities of baryons, neutrons, protons, and $\alpha$ particles respectively, $m_{\rm b}$ is the baryon mass, $k_{\rm B}$ is the Boltzmann constant, $\hbar$ is the reduced Planck's constant, and $Q_{\alpha} \simeq 28.3$\,MeV is the nuclear binding energy of an $\alpha$ particle.

We have modified the Helmholtz EOS to include the effects of nuclear binding energy from $\alpha$ particle dissociation or formation. We do this by modifying the specific internal energy in the EOS as \citep{Fernandez&Metzger13}
\begin{equation}
    e_{\rm int} = e_{\rm int,0} - \frac{Q_{\alpha}}{m_\alpha}X_{\alpha},
\end{equation}
where $e_{\rm int,0}$ is the specific internal energy of the pure nucleon state, $m_{\alpha}$ is the $\alpha$ particle mass, and $X_{\alpha}$ is the mass fraction of $\alpha$ particles. We also include the effects of $\alpha$ particles on the thermodynamic derivatives in the EOS. 

Since $e_{\rm int}$ can be negative for certain combinations of $e_{\rm int,0}$ and $X_{\alpha}$, we take additional care while evolving the internal energy. If the internal energy is negative, the flooring procedures used in the GRMHD code can cause unphysical energy injection. To avoid this, we evolve $e_{\rm int,0} = e_{\rm int} + \frac{Q_{\alpha}}{m_\alpha}X_{\alpha}$, and account for $\frac{Q_{\alpha}}{m_\alpha}X_{\alpha}$ in the EOS subroutines only.  

\subsection{The Neutron Star}
\label{sec:NS}

We choose the NS angular velocity $\Omega_{\star}=0.05\ c/r_{\rm g}$, 
corresponding to a spin period of $P_\star\simeq1.26$ ms, similar to
values expected for rapidly spinning remnants formed in collapsar and
post-merger environments \citep{Mosta+14,Radice+18}. The spacetime is fixed with $a=1/3$ throughout the calculation, but the
imposed rotation of the stellar surface and magnetosphere is initially
set to zero. Beginning at $t=500\,r_{\rm g}/c$, we ramp
$\Omega_\star$ smoothly to its final value over the next
$500\,r_{\rm g}/c$. We assume the NS to be rotating as a solid body. Physically, the NS is likely born differentially rotating, providing a reservoir of free energy that may amplify its magnetic field (e.g., \citealt{Kluzniak&Ruderman98}). However, a variety of instabilities are expected to lead to the redistribution of angular momentum and erase differential rotation on timescales of seconds or less (e.g., \citealt{Margalit+22}). 

We initialize the magnetosphere with the analytic Schwarzschild dipole
solution of \citet{Wasserman1983}, used here as an approximate initial
field configuration in the subsequently evolved Kerr spacetime,
\begin{equation}
    \label{NS_Dipole_Vect}
    A_{\phi}(r,\theta)=\frac{3\mu_{\star} \sin^2 \theta}{2} \left[x^2\ln(1-x^{-1})+x+0.5\right],
\end{equation}
where $x\equiv r/2$, $\mu_{\star}$ is the magnetic dipole moment in units of $r_{\rm g}^3\sqrt{4\pi \rho_{\rm max}c^2}$, and $\rho_{\rm max}$ denotes the peak density in the torus in CGS units (Sec.~\ref{sec:torus}). See \citet{Das2022} for a discussion of accretion onto non-dipolar stellar fields.

As described in Sec.~\ref{sec:suite}, we explore a range of accreting
NS models with surface magnetic field strengths between
$B_{\star}=3\times10^{13}$\,G and $B_{\star}=10^{16}$\,G. We also
simulate an NS with no accretion disk
(model \texttt{NS-B1e15-nodisk}). As a check on the code, we have
verified that the total outflow luminosity of an isolated magnetar
matches the expected force-free spin-down rate, $\dot{E}_{0}$
(Eq.~\eqref{eq:Edotsd}), to within $\lesssim2-3\%$.

\subsection{Initial Torus}
\label{sec:torus}
For the initial torus, we employ the commonly used equilibrium \citet{Fishbone1976} configuration, with an inner edge at $r_{\rm in}=25$\,$r_{\rm g}$, and density maximum at $r_{\rm max}= 47$\,$r_{\rm g}$. The former is larger compared to previous simulations of BH accretion (e.g., \citealt{Siegel&Metzger18} take $r_{\rm in} \approx 4$\,$r_{\rm g}$) to enable a clear separation between the initial torus and the NS magnetosphere.

We set a constant electron fraction $Y_{\rm e}=0.4$ in the initial torus, though our results are not sensitive to this choice because we shall find that weak interactions in the disk rapidly reset $Y_{\rm e}$. To set the initial density ($\rho$) and temperature ($T$) profiles of the torus, we first have to specify the specific enthalpy $h=1+\epsilon/\rho+p/\rho$, and the entropy per baryon. We follow \citet[their Eq.~(3.6)]{Fishbone1976} to obtain the specific enthalpy $h(\rho, T,Y_{\rm e})$, and then pick a constant entropy $s(\rho, T,Y_{\rm e})=5$\,$\rm k_b \ baryon^{-1}$. Although derived assuming a barotropic EOS, the \citet{Fishbone1976} expression for enthalpy is valid here because we take entropy to be constant throughout the initial torus. Because our EOS includes $\alpha$ particles, the internal energy can be negative, and so the torus boundary is not defined by a minimum specific enthalpy of $h_{\rm min}=1$. We instead choose $h_{\rm min}= 0.9939$ to define the torus boundary. We then simultaneously solve enthalpy and entropy equations using a 2D Newton-Raphson method to set the density and temperature at each grid point. These choices of the torus geometry, entropy, and enthalpy yield a peak density of $8.5\times 10^{9}$\,g cm$^{-3}$, and a peak temperature of $1.65\times 10^{10}$\,K. The resulting torus has a total mass of $0.2\,M_{\odot}$.  This mass should not be interpreted as the total mass available in the
global system, particularly in the collapsar application where fallback
from the stellar envelope may continue to replenish the disk for many
seconds. Rather, in our local engine calculation the torus mass is chosen
to provide a finite reservoir that produces the desired hyper-accretion
rate, $\dot{M}\sim 0.01$--$0.1\,M_\odot\,{\rm s}^{-1}$, over the simulated
time interval.

\begin{figure*}
\centering
\safeincludegraphics[width=0.93\textwidth]{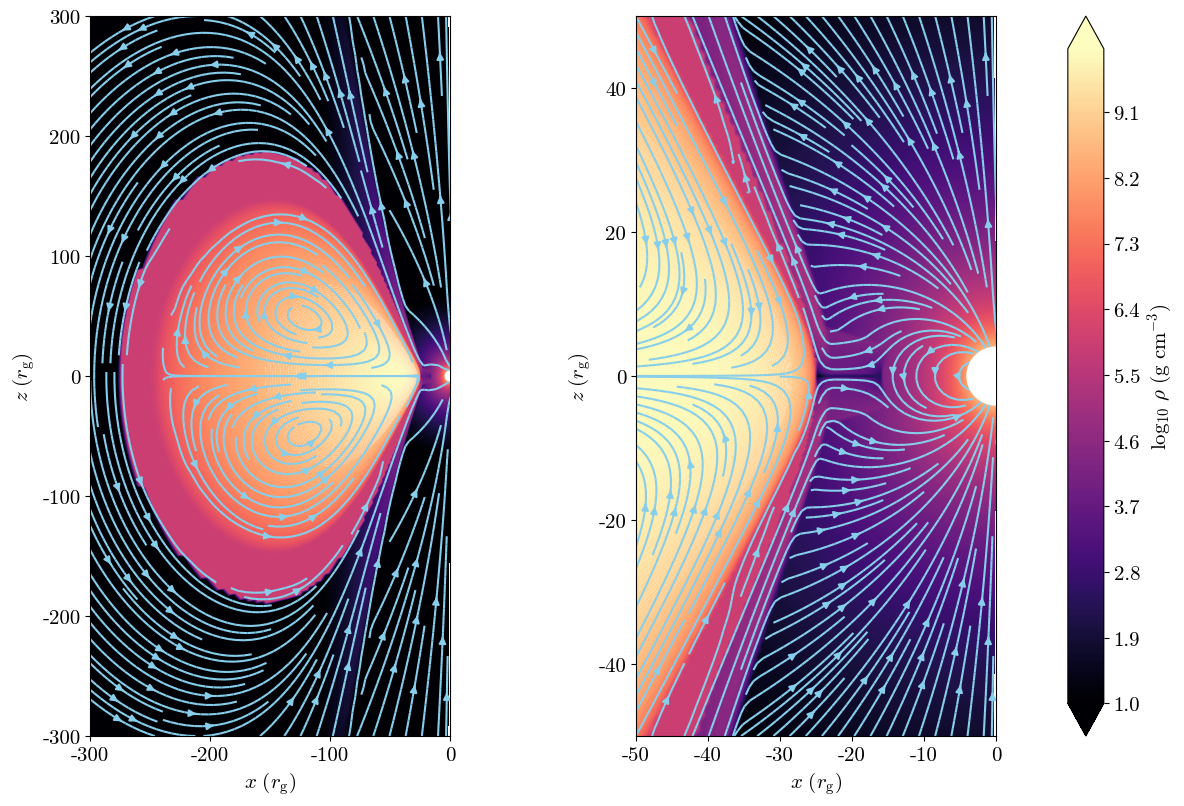}
\caption{Structure of the initial torus at large (left panel) and small (right panel) radial scales in the left half of the $x-z$ plane. Shading denotes contours of the density $\rho$, while the lines with arrows indicate the magnetic field lines. The deformation of the initial dipole magnetic field of the NS to avoid the torus is evident.}
\label{fig:initialtorus}
\end{figure*}

Since the initial torus extends inward to $25\,r_{\rm g}$, we deform the shape of the initial NS magnetosphere (see Fig.~\ref{fig:initialtorus}) to avoid the torus \citep{Parfrey&Tchekhovskoy24}. If we omit this step, the strong NS field is baked into the torus through the initial conditions. The deformation is justified because $r_{\rm in}$ is located outside the eventual light cylinder radius, $R_{\rm LC}$ (as described in Sec.~\ref{sec:NS}, the NS achieves its maximum angular velocity by $t=1000$\,$r_{\rm g}/c$). 

We choose two poloidal loops of opposite polarity across the equator
to minimize the net vertical magnetic flux delivered to the central
object and thereby maintain a SANE rather than MAD accretion state.
This permits a cleaner comparison of the outflows generated by the
stellar dipole across the NS sequence.


We use a purely poloidal vector potential $A_{\phi, \rm torus} \propto {\rm MAX}(\rho-\rho_{\rm cut},0)r^4\sin^{3}(\theta) \cos(\theta)$ for the torus magnetic field, where the cut-off density $\rho_{\rm cut}=10^{-3}\rho_{\rm max}$ limits the magnetic field to only the dense parts of the torus. We normalize the torus magnetic field such that the minimum plasma $\beta=2p/b^2$ (where $b^2$ is the squared magnitude of the fluid-frame magnetic 4-vector in Lorentz-Heaviside units) is 300 in those cells of the torus that have a density of at least $0.05\,\rho_{\rm max}$. We add a random $\sim 4\%$ perturbation to the initial energy-density
profile to seed the turbulence. 

\begin{figure*}
\centering
\safeincludegraphics[width=\textwidth]{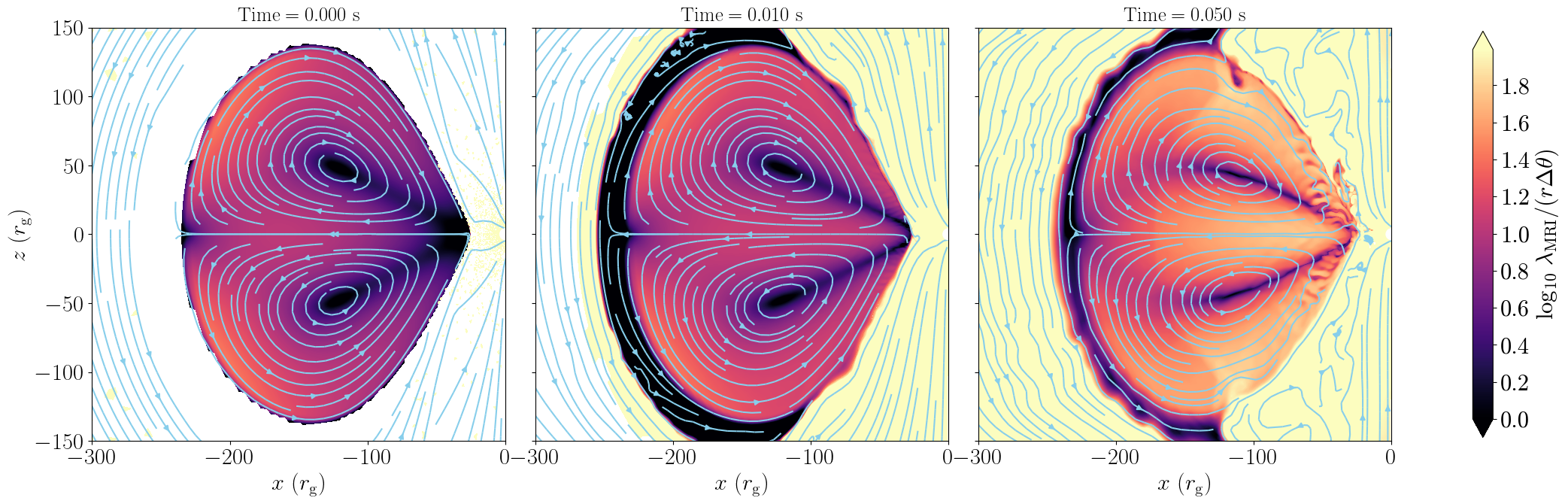}
\caption{Angular MRI quality factor
$Q_{\theta}\equiv\lambda_{\rm MRI}/(r\Delta\theta)$ at three representative
times. The fastest-growing MRI mode is resolved by
$Q_{\theta}\gtrsim10$ throughout most of the dense torus after the
instability develops. Lower values occur primarily near the symmetry
axes, where the imposed initial
poloidal field is weak.}
\label{fig:mri}
\end{figure*}

We assess the numerical resolution of the fastest-growing MRI mode with wavelength \citep{Siegel&Metzger18}
\begin{equation}
    \lambda_{\rm MRI}=\frac{2\pi}{\Omega}\frac{b}{\sqrt{\rho h + b^2}},
\end{equation}
where $h$ is the specific enthalpy, $b=\sqrt{b^2}$ is the fluid-frame magnetic field, and $\Omega=u^{\phi}/u^t$. The quality factor is defined as
\begin{equation}
Q_{\theta} \equiv \frac{\lambda_{\rm MRI}}{r\Delta\theta},
\end{equation}
where $r\Delta\theta$ is the angular grid spacing. Fig.~\ref{fig:mri} shows that the fastest-growing MRI mode is
resolved by $Q_\theta\gtrsim10$ throughout most of the dense torus
after the instability develops. This establishes adequate local
resolution of the linear MRI, although sustained MRI turbulence cannot
be captured indefinitely in axisymmetry.

We note that for the same torus geometry, the density scale and corresponding mass of the torus can be changed by varying its initial entropy. This also allows us to tune the time-averaged mass accretion rate, to values $\dot{M} \sim 0.01-0.1\,M_{\odot}$ s$^{-1}$ in the range expected for collapsar or post-merger disks (Eq.~\ref{eq:Mdot}). The average mass accretion rate can also be changed by varying the distribution of plasma $\beta$ in the initial torus.   

\subsection{Weak Interactions}
\label{sec:weak}

Weak interactions are important in both cooling the accretion flow and changing its electron fraction.  We include the effects of the charged-current URCA pair capture reactions,
\begin{eqnarray}
    e^{-}+p\rightarrow n+\nu_{\rm e} \nonumber \\
    e^{+}+n\rightarrow p+\bar{\nu}_{\rm e}
    \label{eq:URCA}
\end{eqnarray}
which are the dominant neutrino emission channel for the high densities and temperatures of interest.

We calculate the neutrino cooling $Q(\rho,T,Y_{\rm e})$ and lepton-changing rates $R(\rho,T,Y_{\rm e})$ which enter Eqs.~\eqref{energy_eqn}, \eqref{YE_evol_eqn}, arising from the capture reactions for electrons and positrons of arbitrary relativism and degeneracy following \citet{Chen&Beloborodov07}, \citet{Chen2025}.  As our GRMHD simulations do not include neutrino transport, we assume optically thin neutrino cooling.  This approximation is reasonable because the optical depth out of the accretion flow in our simulations to $\nu_e$ and $\bar{\nu}_e$ emitted by the disk is $\lesssim 1$ at all times, as we verify in Appendix \ref{sec:optically_thin}.

Neutrino absorption can also heat the disk surface and modify the electron fraction of the unbound ejecta.  The dominant neutrino irradiation generally originates at small radii,
from the inner accretion flow, a cooling accretion layer on the NS
surface, or the diffusive cooling luminosity of the newly formed
proto-neutron star. Although we neglect neutrino absorption in our fiducial simulations, we assess its influence on the energetics and electron fraction of the disk outflows by including in a subset of models a central isotropic “light-bulb” neutrino source, intended to mimic emission from near the NS surface.  Specifically, we include an additional optically thin heating source term $Q \propto L_{\nu}/r^{2}$ in Eq.~\eqref{energy_eqn} and an additional $Y_{\rm e}-$changing absorption term $R \propto L_{\nu}/r^{2}$ in Eq.~\eqref{YE_evol_eqn} following \citet[their Eq.~(10) and Eq.~(64), respectively]{Qian1996}, assuming equal electron-neutrino and antineutrino luminosities,
$L_{\nu_e}=L_{\bar\nu_e}$, and
two values of the total electron-flavor luminosity,
$L_{\nu_{\rm e}}+L_{\bar{\nu}_{\rm e}}
=2\times10^{51}$ and $2\times10^{52}\,{\rm erg\,s^{-1}}$, spanning the expected range given the ages and accretion rates of the NSs of interest (see Fig.~\ref{fig:neutrino_lumtemp} and Paper II for further discussion). In calculating the absorption cross sections, we assume mean neutrino energies $\epsilon_{\nu_{\rm e}}=\epsilon_{\bar{\nu}_{\rm e}}=10$\,MeV, typical of proto-neutron stars or post-merger remnants several seconds in their cooling evolution (e.g., \citealt{Pons+99,Dessart+09}). Because the electron-fraction-changing absorption rates scale
approximately as $L_\nu\epsilon_\nu$, the adopted luminosity and mean
energy should be viewed jointly: a larger luminosity with a lower mean
energy can produce a similar degree of weak processing. Because the effects of neutrino irradiation do not appreciably alter the disk-magnetosphere interaction and are mostly relevant to the composition and nucleosynthesis of the ejecta, we defer a detailed analysis of the irradiated models to Paper~II.

\subsection{Model suite}
\label{sec:suite}
Our suite of NS and BH simulations is summarized in Table \ref{tab:models}.  All models assume a compact object of mass $M_{\star} = 2\,{M}_{\odot}$ and the same angular momentum, corresponding to a Kerr parameter $a = 1/3$ and, in the NS models, to a spin period $P_{\star} = 1.26$ ms.  Other key radii include the innermost stable circular orbit (ISCO), $R_{\rm isco} \simeq 4.9 \,r_{\rm g}$; light cylinder radius $R_{\rm LC}=20\, r_{\rm g}$ (Eq.~\eqref{eq:RL}), and corotation radius $R_{\rm c} = 7.3\, r_{\rm g}$ (Eq.~\eqref{eq:Rcor}).

Given our assumed initial torus mass ($M_{\rm d} = 0.2M_{\odot}$) and its characteristic outer radius, $r_{\rm max} = 47 r_{\rm g}$, Eq.~\eqref{eq:Mdot} predicts an inflow rate $\dot{M} \gtrsim 0.1 \, {\rm M}_{\odot} \ \rm $s$^{-1}$ (for $h/r \sim 0.5$ at the density maximum), for values of the effective viscosity $\alpha \sim 10^{-2}$ similar to those generated by the MRI. This exceeds the critical accretion rate $\dot{M}_{\rm ign} \sim 10^{-2}M_{\odot}$ s$^{-1}$ (Eq.~\eqref{eq:Mdotign}) necessary for the disk to become neutron-rich at radii $\lesssim r_{\rm max}$ on the inflow timescale.  Thus, we expect the inflowing matter to be neutron-rich with $Y_{\rm e} \sim 0.1$ by the time it reaches the magnetosphere, regardless of its initial composition.

The accreting NS sequence spans equatorial surface dipole strengths
from $B_\star=3\times10^{13}$ to $10^{16}\,{\rm G}$. In
\texttt{NS-B3e13} and \texttt{NS-B3e14}, the magnetosphere is
compressed to the stellar surface. Model \texttt{NS-B1e15} remains
nominally within the crushed-magnetosphere regime, although its
compressed field is already dynamically important near the surface.

At larger field strengths, the disk is increasingly truncated outside
the star. Model \texttt{NS-B3e15} lies in the magnetically channeled
accretion regime, with
$R_{\star}<R_{\rm m}<R_{\rm c}$. Model \texttt{NS-B4e15} lies close to
corotation, $R_{\rm m}\simeq R_{\rm c}$, and therefore near the
transition between accretion-dominated spin-up and magnetically
dominated spin-down. Finally, model \texttt{NS-B1e16} lies securely in
the propeller regime, $R_{\rm m}>R_{\rm c}$, where centrifugal coupling
to the rotating magnetosphere suppresses direct accretion.
This expanded sequence allows us to follow the continuous progression
from an accretion-dominated crushed magnetosphere, through channeled
accretion and approximate spin equilibrium, to propeller-driven
spin-down.

To explore the influence of neutrino absorption
(Sec.~\ref{sec:weak}), we perform irradiated counterparts of the
black-hole model, the near-corotation model \texttt{NS-B4e15}, and the
propeller model \texttt{NS-B1e16}. These models are denoted
\texttt{BH$\nu$2e51}, \texttt{NS-B4e15$\nu$2e51},
\texttt{NS-B1e16$\nu$2e51}, and
\texttt{NS-B1e16$\nu$2e52}, where the suffix gives the imposed total
electron-flavor neutrino luminosity in ${\rm erg\,s^{-1}}$. We also
perform a spinning NS calculation without an accretion disk,
\texttt{NS-B1e15-nodisk}, to provide a reference for isolated
magnetar spin-down. The simulations are run for a total duration of
$0.9$ seconds ($\sim 9\times10^{4}\,r_{\rm g}/c$), sufficient to capture the
disk--magnetosphere interaction over several local inflow times. All time averages presented in this paper are taken from
$0.4$ to $0.9\,{\rm s}$ after the start of the simulation,
by which time the electron fraction has decreased sufficiently
from its initial value in the torus and the inflow has reached
an approximately steady composition.

\begin{deluxetable}{lccccccccccc}
\tablecaption{Summary of Simulations\label{tab:models}}
\tablehead{
Model$^{(1)}$   & $B_{\star}^{(2)}$ & $\langle R^{\rm e_1}_{\rm m} \rangle ^{(3)}$ & $\langle R^{\rm e_2}_{\rm m}\rangle ^{(3)}$ & $\dot{M_{\star}} ^{(4)}$ \\
 &    (G) & ($r_{\rm g}$) & ($r_{\rm g}$) & ($M_{\odot} \ s^{-1}$) 
}
\startdata
\texttt{BH}                 & -- & -- & --  & $1.9\times 10^{-2}$              \\
\texttt{BH$\nu$2e51}                 & -- & -- & --  & $1.5\times 10^{-2}$              \\
\texttt{NS-B3e13}          & $3\times 10^{13}$   & $4.0\pm 0.0$ & $4.0\pm 0.01$ & $1.2\times 10^{-2}$ \\
\texttt{NS-B3e14}          & $3\times 10^{14}$   & $4.0\pm 0.0$ & $4.0\pm 0.01$ & $9.0\times 10^{-3}$ \\
\texttt{NS-B1e15-nodisk}   & $10^{15}$          & --  & -- & -- \\
\texttt{NS-B1e15}          & $10^{15}$   & $4.3\pm 0.4$ & $4.1\pm 0.1$ & $10^{-2}$ \\
\texttt{NS-B3e15}          & $3\times 10^{15}$   & $5.9\pm 0.5$ & $5.1\pm 0.7$ & $9.0\times 10^{-3}$ \\
\texttt{NS-B4e15}          & $4\times 10^{15}$  & $6.5\pm 0.5$ & $6.0\pm 0.7$ & $8.0\times 10^{-3}$ \\
\texttt{NS-B4e15$\nu$2e51}          & $4\times 10^{15}$  & $6.8\pm 0.5$ & $6.4\pm 0.7$ & $7.0\times 10^{-3}$ \\
\texttt{NS-B1e16}          & $10^{16}$  & $10.8\pm 2.3$ & $10.8\pm 2.4$ & $4.0\times 10^{-3}$ \\
\texttt{NS-B1e16$\nu$2e51} & $10^{16}$  & $12.8\pm 1.3$ & $12.8\pm 1.4$ & $3.0\times 10^{-3}$ \\
\texttt{NS-B1e16$\nu$2e52} & $10^{16}$  & $17.4\pm 2.7$ & $17.4\pm 2.3$ & $10^{-4}$ \\
\enddata

\tablecomments{All models assume a compact object mass $M_{\star} = 2\,M_{\odot}$ and Kerr spin parameter $a = 1/3$, corresponding to a NS spin period $P_{\star} = 1.26$ ms, and an initial torus mass $M_{\rm d}=0.2\,{M}_\odot$ and initial entropy $s = 5\, \rm k_b$ baryon$^{-1}$.  Critical radii include: the NS surface $R_{\star} = 4\, r_{\rm g}$, ISCO $R_{\rm isco} = 4.9\,r_{\rm g}$, corotation $R_{\rm c} = 7.3\, r_{\rm g}$ (Eq.~\eqref{eq:Rcor}), and light cylinder $R_{\rm LC} = 20\,r_{\rm g}$ (Eq.~\eqref{eq:RL}). For all the NS models, we report the NS magnetic field in the fluid frame. The time-averaged quantities have been averaged from 0.4\,s to 0.9\,s after the start of the simulation, to allow sufficient time for $Y_{\rm e}$ to decrease from its initial value in the torus. \\
$^{(1)}$Model identifier: Neutron-star models are named according to
their equatorial surface dipole field strength, while the suffix
$\nu$ followed by a luminosity identifies simulations with an imposed
neutrino light bulb.\\
$^{(2)}$Surface equatorial magnetic field strength of the NS in fluid frame; this is not applicable for BH accretors.\\
$^{(3)}$Angle- and time-averaged magnetosphere (Alfv\'en) radius over zones within $\pi/24$ of the equator for the accreting NS models (Fig.~\ref{fig:MagRad}), calculated using two different methods (Sec.~\ref{sec:mag_structure}). Values are reported in the format $\rm average \pm standard \ deviation$.\\
$^{(4)}$Time-averaged mass accretion rate, as measured through the inner boundary of the simulation. 
}
\end{deluxetable}

\begin{figure*}
\centering
\safeincludegraphics[width=0.95\textwidth]{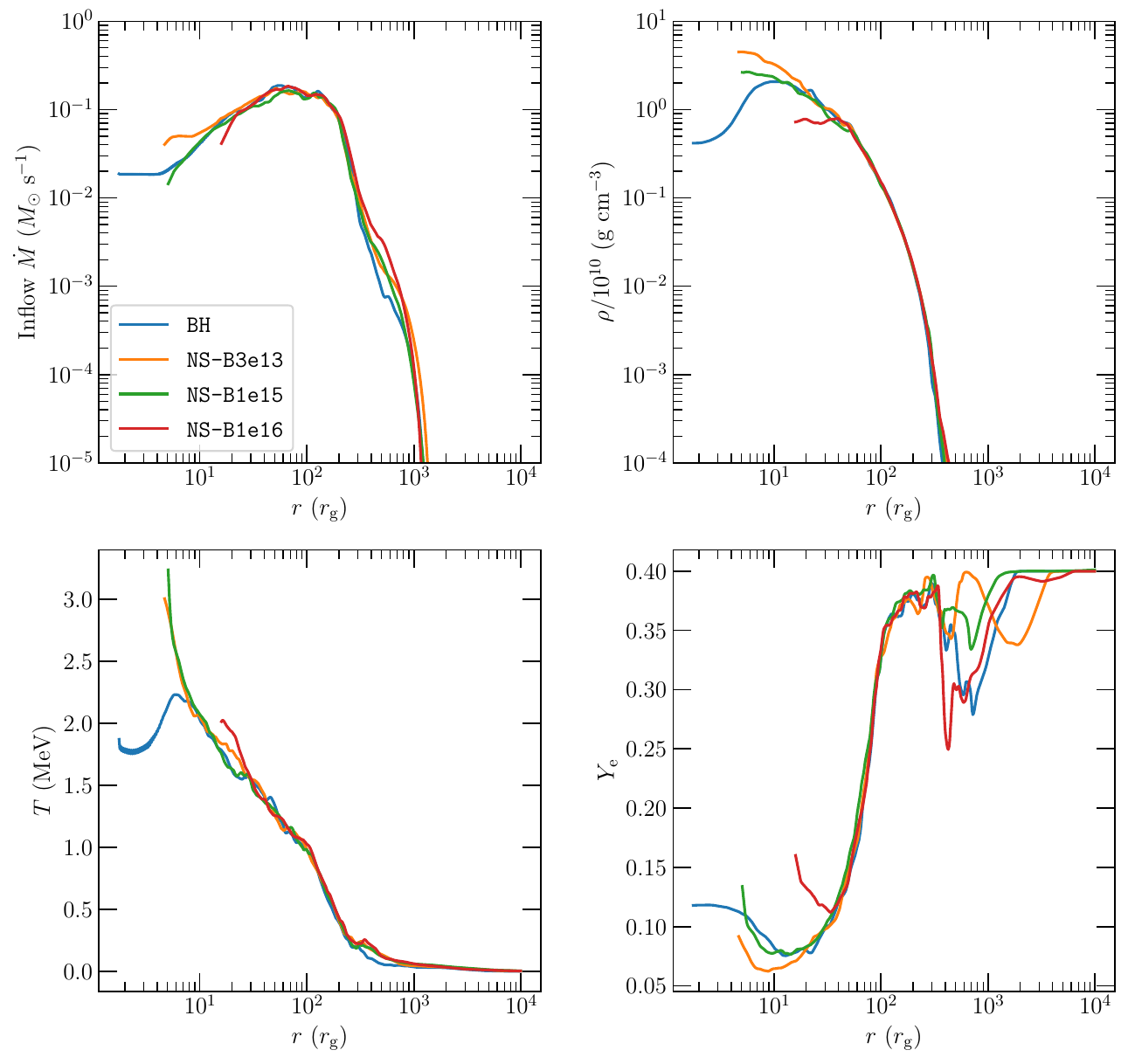}
\caption{Time-averaged (from 0.4\,s to 0.9\,s after the start of the simulation) radial profiles of mass inflow rate, density, temperature, and electron fraction.  The mass inflow rate is measured through spherical shells while the latter three are measured in the equatorial plane.  The drop in $\dot{M}$ inside $r \lesssim 40\, r_{\rm g}$ demonstrates the development of strong outflows. We terminate the NS profiles at the average magnetosphere radius (or the NS radius, whichever is larger).}
\label{fig:radial_profiles}
\end{figure*}

\section{Results}
\label{sec:results}
\subsection{Disk--Magnetosphere Interaction}
\label{sec:mag_structure}

The simulations are designed so that each compact object is fed by
essentially the same accretion flow at large radii. This is shown in
Fig.~\ref{fig:radial_profiles}, whose time-averaged radial profiles
demonstrate that the net mass inflow rate supplied by the torus is
nearly identical across the BH and NS models, with an outer feeding
rate $\dot{M}_{\rm d}\simeq 0.2\,M_\odot\,{\rm s^{-1}}$. The
thermodynamic structure of the disk is also similar at large radii:
the density and temperature profiles follow the expected behavior of
a hot, neutrino-cooled accretion flow, and the gas neutronizes to
$Y_{\rm e}\simeq 0.1-0.15$ near the density maximum.

The models therefore differ primarily in what happens once this common
inflow reaches the central object. For the BH, and for sufficiently
weak neutron-star dipole fields, the gas can continue inward with
little interruption. For stronger $B_\star$, however, the stellar magnetic field halts the
disk outside the stellar surface, and the subsequent evolution depends
on the location of the magnetospheric radius $R_{\rm m}$ relative to the
stellar radius $R_\star$ and the corotation radius $R_{\rm c}$.

For stronger $B_\star$, however, the stellar
magnetic field halts the disk outside the surface, and the subsequent
evolution depends on the location of the magnetospheric radius
$R_{\rm m}$ relative to the stellar radius $R_\star$ and the corotation
radius $R_{\rm c}$. Thus, the same large-scale feeding rate can lead to
a crushed magnetosphere, magnetically channeled accretion, or a
centrifugal propeller, depending on $B_\star$.

\begin{figure*}
\centering
\safeincludegraphics[width=0.95\textwidth]{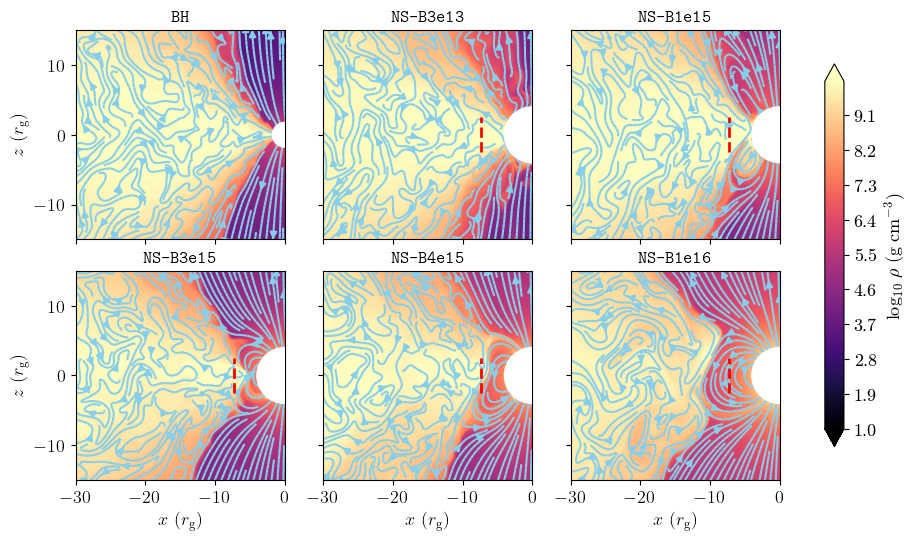}
\caption{Inner accretion-flow structure for the black-hole model and for neutron-star models spanning different disk--magnetosphere regimes at time $t = 0.4$\,s after the onset of the simulation.
At fixed outer mass supply, increasing the neutron-star dipole field moves the magnetospheric radius $R_{\rm m}$ from a crushed state, $R_{\rm m} \lesssim R_\star$, to magnetically-channeled accretion, $R_\star \lesssim R_{\rm m} \lesssim R_{\rm c}$, and finally to a centrifugal propeller, $R_{\rm m} \gtrsim R_{\rm c}$. Color denotes gas density, arrowed curves trace magnetic field lines, and vertical red dashed lines mark the corotation radius $R_{\rm c}$ in the neutron-star panels.}
\label{fig:mag_structure_4panel}
\end{figure*}

\begin{figure*}
\centering
\safeincludegraphics[width=\textwidth]{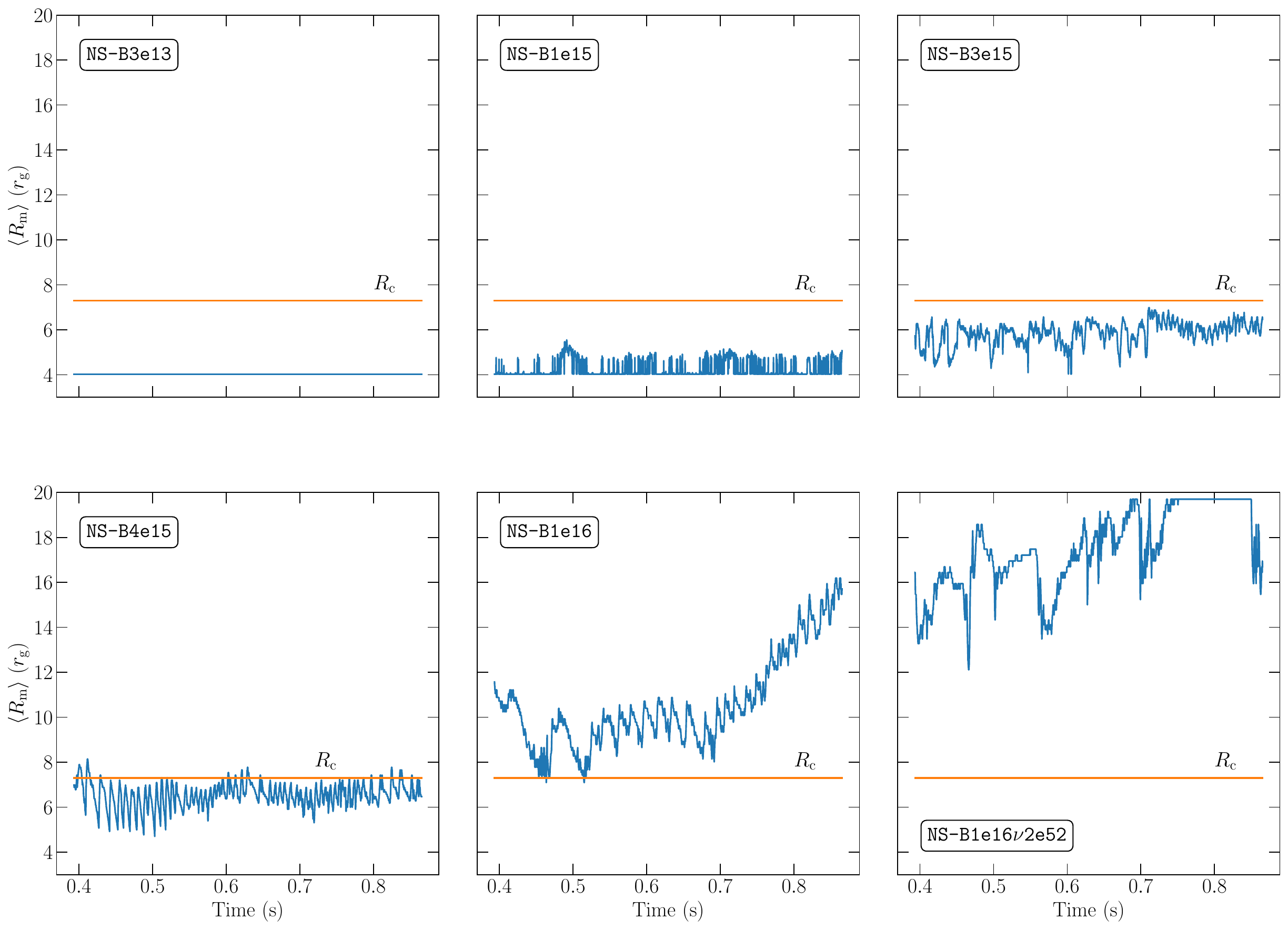}
\caption{Angle-averaged magnetosphere radius in the zones within $\pi/24$ of the equator as a function of time, for models spanning different NS magnetic fields $B_{\star}$. The horizontal orange lines denote the corotation radius, $R_{\rm c}$.  We show results starting at $t = 0.4$ s to capture the approximately steady state after $Y_{\rm e}$ has decreased from its initial value. }
\label{fig:MagRad}
\end{figure*}

\begin{figure}
\centering
\safeincludegraphics[width=\linewidth]{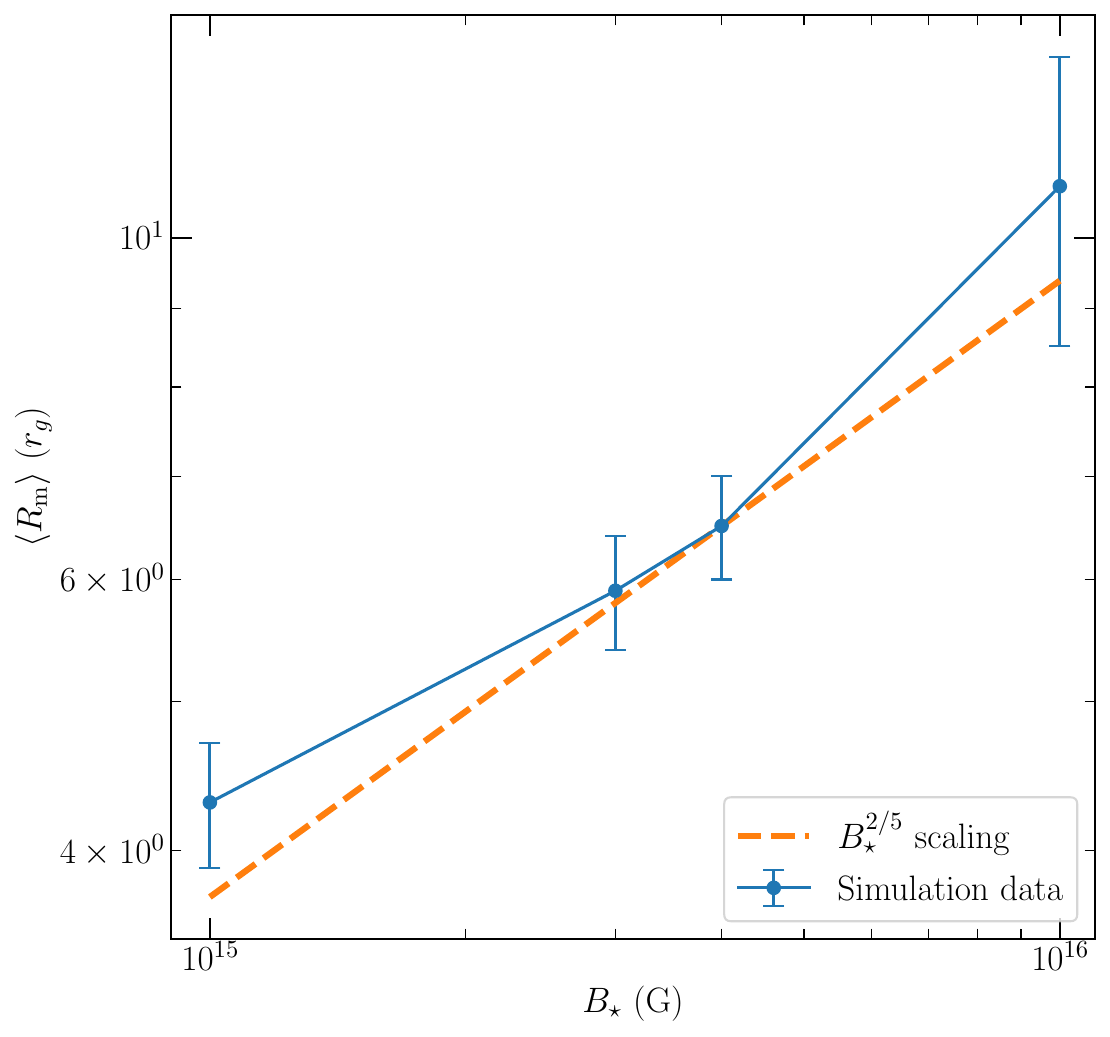}
\caption{Time- and angle-averaged magnetosphere radius from the simulations as a function of the NS magnetic field. Comparison with the scaling $R_{\rm m}\propto B_{\star}^{2/5}$ \citep{Kulkarni2013} shows reasonable agreement.}
\label{fig:MagRadScaling}
\end{figure}

\begin{figure*}
\centering
\safeincludegraphics[width=\textwidth]{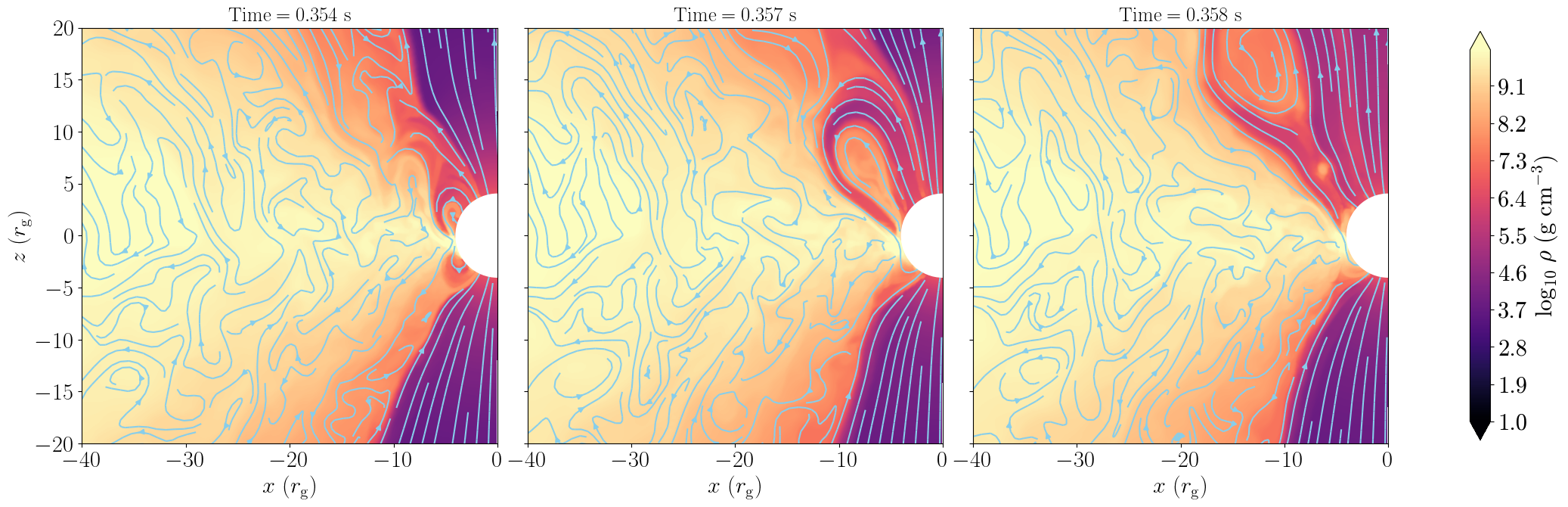}
\caption{Sequence of snapshots of the model \texttt{NS-B1e15}, showing a
plasmoid eruption driven by magnetic reconnection when the
magnetosphere is pushed close to the NS surface.}
\label{fig:plasmoid}
\end{figure*}

\begin{figure*}
\centering
\safeincludegraphics[width=\textwidth]{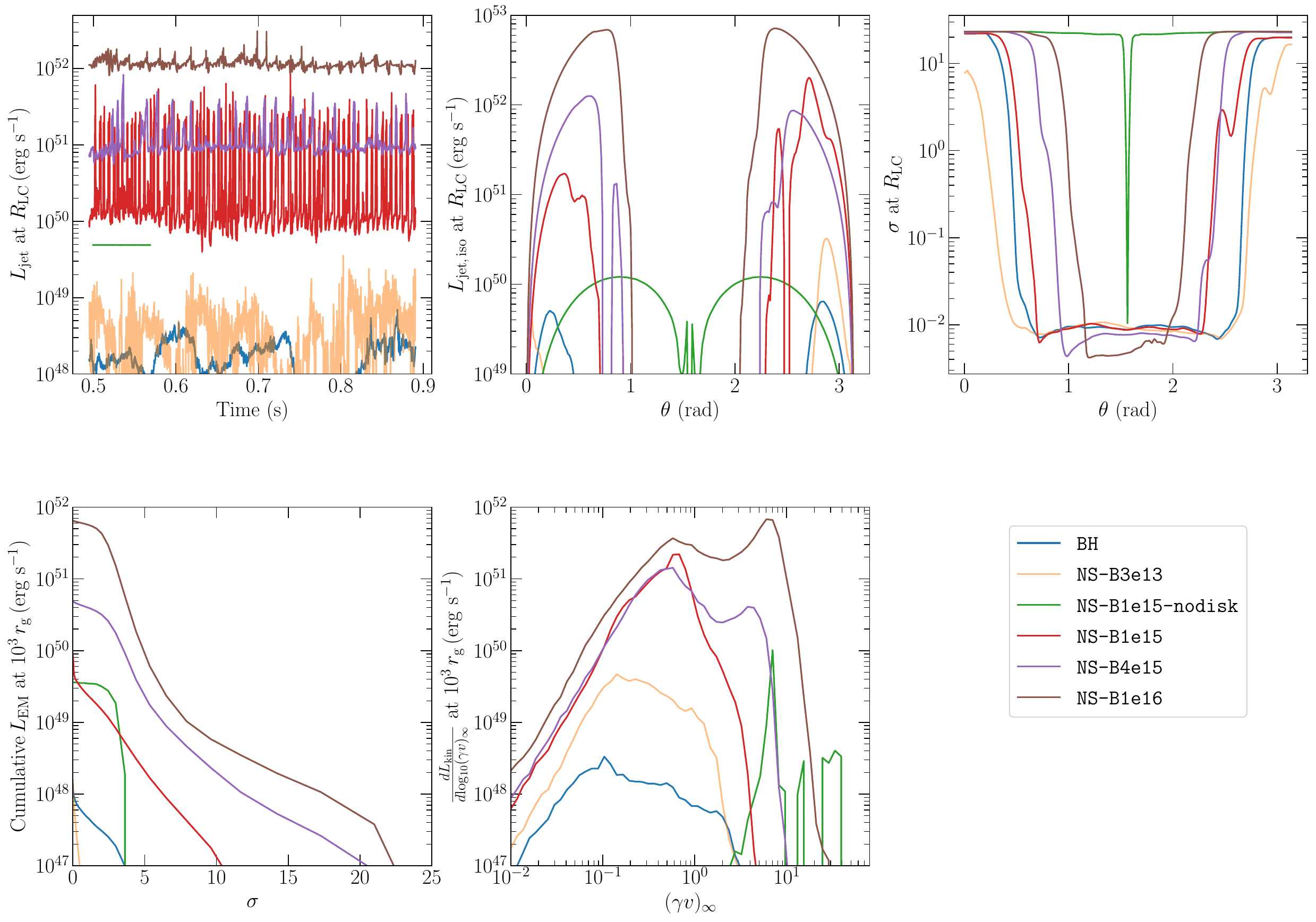}
\caption{
Outflow energetics, magnetization, and velocity structure across models.
The upper-left panel shows the time-dependent relativistic jet power, defined as the Poynting flux in $\sigma > 1$ material measured at $R_{\rm LC} = 20\,r_{\rm g}$. The upper-middle and upper-right panels show the time-averaged angular profiles of the isotropic-equivalent jet luminosity, $4\pi\, dL_{\rm EM,\sigma>1}/d\Omega$, and magnetization $\sigma$, respectively.
The lower-left panel shows the cumulative electromagnetic luminosity above a given magnetization, while the lower-middle panel shows the kinetic luminosity as a function of asymptotic radial four-velocity, $(\gamma v)_\infty$.
Together, these diagnostics show how the disk--magnetosphere interaction redistributes the outflow power between magnetically dominated jets, lower-$\sigma$ winds, and faster ejecta.
The absolute normalization of $\sigma$ in the polar jet depends on the
imposed floor density, which stands in for unresolved baryon loading of
the proto-magnetar jet by neutrino heating from the stellar atmosphere. The shown distribution of $\sigma$ for $\sigma \gtrsim 5$ is thus dependent on the assumed magnetization of the magnetar jet.
}
\label{fig:sigma_jet_power_theta}
\end{figure*}

Figure~\ref{fig:mag_structure_4panel} illustrates this progression in
snapshots of the inner accretion flow. In the \texttt{BH} model, the
inner accretion flow is turbulent, but little mass is launched from the
inner disk at $r\lesssim15\,r_{\rm g}$, and most of the gas that reaches
these radii falls through the inner boundary (see also the nearly
radially constant $\dot{M}$ in Fig.~\ref{fig:radial_profiles}). In the
lowest-field NS model, \texttt{NS-B3e13}, the stellar field has little
effect on the inflow and the magnetosphere is compressed completely
against the stellar surface. The magnetosphere likewise remains
nominally crushed in \texttt{NS-B1e15}, although the stronger compressed
field produces appreciable structure close to the surface. At larger
field strengths, the disk is increasingly truncated outside the star:
\texttt{NS-B3e15} exhibits magnetically channeled accretion,
\texttt{NS-B4e15} lies close to corotation, and \texttt{NS-B1e16}
lies securely in the propeller regime, where centrifugal coupling to
the rotating magnetosphere suppresses direct accretion onto
the star.

The remaining panels of Fig.~\ref{fig:radial_profiles} show that the
thermodynamic structure of the disk is broadly similar among the models
outside the immediate vicinity of the compact object, and agrees well
with one-dimensional models of neutrino-cooled disks
\citep{Chen&Beloborodov07}. The disk neutronizes rapidly, reaching
$Y_{\rm e}\simeq 0.1-0.15$ near the density maximum of the outer torus.
The main differences appear only at small radii, where the flow begins
to interact directly with the inner boundary or the stellar
magnetosphere. In the lowest-field model \texttt{NS-B3e13}, the
profiles remain close to those of the \texttt{BH} model because the
stellar field is too weak to maintain a distinct magnetospheric cavity.
In \texttt{NS-B1e15}, density and temperature enhancements appear near
the stellar surface as the compressed magnetosphere begins to modify
the inflow. In the propeller model \texttt{NS-B1e16}, the disk is
truncated at $r\simeq10-15\,r_{\rm g}$.

To quantify these regimes, Fig.~\ref{fig:MagRad} shows the time
evolution of the angle-averaged magnetospheric radius
$\langle R_{\rm m}\rangle$. We estimate this radius using two
diagnostics, following \citet{Zhu+24}. First, moving inward from the
initial torus edge at $r_{\rm in}=25\,r_{\rm g}$, we identify the
radius at which the magnetization, averaged over zones within
$\pi/24$ of the equator, first exceeds unity:
$\langle\sigma\rangle=\langle b^2/(h\rho)\rangle>1$
\citep{Parfrey&Tchekhovskoy24}.

As a second estimate, we identify the radius at which the magnetic
energy density,
$E_{\rm b}=b^t b_t-b^2u^tu_t-b^2/2$, first exceeds the kinetic
energy density, $E_{\rm kin}=-\rho(u_t+1)$, after summing both
quantities over the same equatorial angular range.

Fig.~\ref{fig:MagRadScaling} shows the time-averaged value of
$\langle R_{\rm m}\rangle$ as a function of $B_\star$. The measured
scaling agrees reasonably well with
$R_{\rm m}\propto B_\star^{2/5}$, as expected for a disk-compressed
magnetosphere \citep{Kulkarni2013}, rather than the simpler spherical
Alfv\'en scaling $R_{\rm m}\propto B_\star^{4/7}$ from Eq.~\eqref{eq:Rm}.

When the magnetosphere is pushed close to the NS surface,
differential rotation between the star and disk twists the connected
field lines, building up toroidal magnetic flux. The field inflates
outward into a helmet-streamer-like configuration
\citep{Aly1990,LyndenBell&Boily94,Uzdensky2004}, which becomes
unstable to reconnection and plasmoid ejection
\citep{Zanni+13,Parfrey+17}. Fig.~\ref{fig:plasmoid} shows a
representative eruption. Such events recur while the magnetosphere
remains compressed near the stellar surface and produce substantial
jet variability (Sec.~\ref{sec:jet}).

\begin{deluxetable*}{lcccccccccccccc}
\tablecaption{Outflow and Jet Luminosities, Torques \label{tab:LumJet}}
\tablehead{
Model     & $L_{{\rm jet}}^{(1)}$ & $L_{{\rm jet}}/\dot{E}_0^{(2)}$ &  $\tau/|\tau_0|^{(3)}$ & $J/\tau^{(4)}$ & $L_{{\rm EM},\sigma>1 }^{(5)}$ & $L_{\rm EM, tot}^{(6)}$  & $L_{\rm kin}^{(7)}$  \\
 &    ($\rm erg \ $s$^{-1}$) &  & & (s) & ($\rm erg \ $s$^{-1}$) & ($\rm erg \ $s$^{-1}$) &($\rm erg \ $s$^{-1}$)
}
\startdata
\texttt{BH}           &  $(2.2,1.4)\times 10^{48}$ & -- & -- & $(+)11$ & $(4.8,6.0)\times 10^{47}$ & $(2.6,1.5)\times 10^{48}$  & $(1.3,1.8)\times 10^{48}$ \\
\texttt{BH$\nu$2e51}           &  $(2.9,1.2)\times 10^{48}$ & -- & -- & $(+)14$ & $(6.8,6.0)\times 10^{47}$ & $(3.2,1.3)\times 10^{48}$  & $(4.0,3.3)\times 10^{48}$ \\
\texttt{NS-B3e13}           & $(4.1,3.5)\times 10^{48}$ & $(2.1,1.8)\times 10^{2}$ & $(2.6,0.9)\times 10^{5}$ & $(+)8$ & $(1.0,8.1)\times 10^{46}$ & $(3.1,1.3)\times 10^{48}$  & $(3.9,1.7)\times 10^{49}$  \\
\texttt{NS-B3e14}           & $(1.2,1.1)\times 10^{50}$ & $(62.5,57.3)$ & $(1.2,0.8)\times 10^{3}$ & $(+)16$ & $(1.0,4.6)\times 10^{47}$ & $(5.9,3.1)\times 10^{49}$  & $(5.9,2.2)\times 10^{50}$  \\
\texttt{NS-B1e15-nodisk}      & $(4.9,0.006)\times 10^{49}$ & $(1.0,10^{-3})$ & $(-1.0,10^{-3})$ & $(-)1250$ & $(3.5,0.02)\times 10^{49}$ & $(3.6,0.006)\times 10^{49}$  & $(9.0,0.08)\times 10^{48}$   \\
\texttt{NS-B1e15}           & $(5.8,8.4)\times 10^{50}$ & $(11.4,16.7)$ & $(19.5,13.1)$ & $(+)43$ & $(2.4,1.2)\times 10^{49}$ & $(9.4,3.0)\times 10^{49}$  & $(1.0,0.3)\times 10^{51}$  \\
\texttt{NS-B3e15}        & $(9.0,20.4)\times 10^{50}$& $(4.5,10.2)$ & $(3.0,5.0)$ & $(+)69$    & $(1.7,0.4)\times 10^{50}$ & $(2.6,0.6)\times 10^{50}$  & $(1.1,0.7)\times 10^{51}$\\
\texttt{NS-B4e15}       &  $(1.1,0.7)\times 10^{51}$ & $(2.5,1.6)$ & $(-0.4,4.4)$ & $(-)339$  & $(4.0,0.7)\times 10^{50}$ & $(5.0,1.0)\times 10^{50}$  & $(1.0,0.7)\times 10^{51}$\\
\texttt{NS-B4e15$\nu$2e51}       &  $(1.0,0.6)\times 10^{51}$ & $(2.4,1.4)$ & $(-1.4,4.2)$ & $(-)92$  & $(3.8,0.7)\times 10^{50}$ & $(4.8,0.9)\times 10^{50}$  & $(9.5,0.4)\times 10^{50}$\\
\texttt{NS-B1e16}     & $(1.2,0.2)\times 10^{52}$ & $(2.3,0.3)$ & $(-2.5,0.7)$ & $(-)5$ & $(5.7,0.9)\times 10^{51}$ & $(6.5,1.0)\times 10^{51}$  & $(5.2,1.5)\times 10^{51}$\\
\texttt{NS-B1e16$\nu$2e51}     &  $(1.2,0.1)\times 10^{52}$ & $(2.4,0.3)$ & $(-2.3,0.6)$ & $(-)5$ & $(5.9,1.0)\times 10^{51}$ & $(6.6,1.0)\times 10^{51}$  & $(5.3,1.5)\times 10^{51}$    \\
\texttt{NS-B1e16$\nu$2e52}    & $(7.4,1.1)\times 10^{51}$ & $(1.5,0.2)$ & $(-1.1,0.3)$ & $(-)11$  & $(4.5,0.6)\times 10^{51}$ & $(5.0,0.7)\times 10^{51}$  & $(2.5,1.0)\times 10^{51}$ \\
\enddata

\tablecomments{With the exception of the spin-down time, all values are presented in the format (average, standard deviation). For all the NS models, we report the NS magnetic field in the fluid frame. The time-averaged quantities have been averaged from 0.4\,s to 0.9\,s after the start of the simulation, to allow sufficient time for $Y_{\rm e}$ to decrease from its initial value in the torus. For the final three columns, we measure the outflow luminosities at
$r=1000\,r_{\rm g}$, sufficiently far from the compact object while
remaining well inside the outer boundary. Over the range
$r=300$--$2000\,r_{\rm g}$, $L_{\rm EM,tot}$ generally decreases
outward while $L_{\rm kin}$ increases, indicating conversion of
electromagnetic and residual thermal energy into bulk kinetic energy.
The individual luminosities vary by factors of at most
$\sim1.5$--$2$ across this radial range, so measurements at
$1000\,r_{\rm g}$ are adequate for the comparisons made here.   \\
$^{(1)}$ Time-averaged relativistic jet power measured through a spherical surface at the light cylinder radius 20\,$r_{\rm g}$.\\
$^{(2)}$ Relativistic jet power relative to the corresponding power for an isolated NS (Eq.~\eqref{eq:Edotsd}).\\
$^{(3)}$ Time-averaged torque on the NS relative to the magnitude of the corresponding torque on an isolated NS (the torque on an isolated NS is always negative, but we consider only the magnitude here when reporting the relative values). A negative average value indicates spin-down, while a positive average value indicates spin-up. \\
$^{(4)}$ Time-averaged spin-up/spin-down time $J/\tau$, where $J=\frac{2}{5}M_{\star}R_{\star}^2\Omega_{\star}$ is the angular momentum of the NS. Negative in the parentheses indicates spin-down, while positive indicates spin-up. \\
$^{(5)}$ Time-averaged electromagnetic power contained in $\sigma>1$ measured through a spherical surface of radius $1000\,r_{\rm g}$.\\
$^{(6)}$ Time-averaged total electromagnetic power measured through a spherical surface of radius $1000\,r_{\rm g}$.\\
$^{(7)}$ Time-averaged total kinetic (excluding rest mass) power carried by matter, measured through a spherical surface of radius $1000\,r_{\rm g}$.\\
}
\end{deluxetable*}

\subsection{Energetics of the outflows and the polar jet}
\label{sec:jet}

We next quantify how the disk--magnetosphere interaction modifies the power, magnetization, and velocity structure of the outflows. The electromagnetic (EM) luminosity through a spherical surface is
\begin{equation}
    \label{eqn:jet_power}
    L_{\rm EM} = -\int \int (b^2 u^{r}u_{t} - b^r b_t)\sqrt{-g}\,d\theta \, d\phi.
\end{equation}
We also compute the bulk kinetic-energy luminosity of the matter
outflow, excluding rest-mass energy,
\begin{equation}
    \label{eqn:kin_power}
    L_{\rm kin} =
    \int \int \rho(-u_t - 1)u^r \sqrt{-g}\,d\theta \, d\phi .
\end{equation}
We define the relativistic jet power, $L_{\rm jet}$, as the portion of $L_{\rm EM}$ carried by magnetically dominated material with $\sigma>1$, measured at $r=20\,r_{\rm g}$, corresponding to the equatorial light-cylinder radius. The upper-left panel of 
Fig.~\ref{fig:sigma_jet_power_theta} shows the time evolution of $L_{\rm jet}$, while Table~\ref{tab:LumJet} reports the time-averaged values and rms variability.  These diagnostics separate the relativistic, magnetically dominated
polar component from the broader baryon-loaded disk/propeller wind,
and quantify the electromagnetic and bulk kinetic contributions to
the outflow power. 

Accretion substantially enhances the jet power relative to isolated
magnetar spin-down, although the enhancement factor depends strongly
on field strength. In the very weak-field models
\texttt{NS-B3e13} and \texttt{NS-B3e14}, the isolated-dipole power is
itself extremely small, so even a weak disk-associated polar outflow
corresponds to a large formal value of
$L_{\rm jet}/\dot{E}_{0}$. For the magnetar-strength models most
relevant to GRB engines, \texttt{NS-B1e15} through
\texttt{NS-B1e16}, the mean jet luminosity exceeds the corresponding
isolated-dipole value by factors of approximately $2$--$11$
(Table~\ref{tab:LumJet}). This enhancement is the same open-flux effect
estimated in Sec.~\ref{sec:analytic}: once the disk pushes the
magnetospheric radius inside the light cylinder,
$R_{\rm m}<R_{\rm LC}$, additional stellar magnetic flux is opened and
the spin-down power rises above the isolated-dipole value.
The measured enhancement is somewhat
smaller than the simple estimate
$\dot{E}/\dot{E}_{0}\simeq(R_{\rm LC}/R_{\rm m})^{2}$
(Eq.~\eqref{eq:Edot}), but remains consistent at the
order-of-magnitude level. Even in the propeller model
\texttt{NS-B1e16}, where direct accretion onto the stellar surface is
suppressed, the jet luminosity remains enhanced by a factor
$\simeq2$ relative to isolated dipole spin-down. Thus, the usual
isolated-dipole estimate underpredicts the relativistic power of an
accreting proto-magnetar.

The disk--magnetosphere interaction also makes the jet highly
time-variable. This variability is most pronounced in the models where the
magnetosphere is pushed close to the NS surface and undergoes repeated
reconnection-driven plasmoid eruptions
(Fig.~\ref{fig:plasmoid}). These events produce large spikes in
$L_{\rm jet}$, in sharp contrast to the disk-free magnetar model
\texttt{NS-B1e15-nodisk}, whose outflow is smooth and nearly steady.
The variability is reduced in the propeller model \texttt{NS-B1e16},
for which the magnetosphere remains farther from the stellar surface
and direct accretion is less frequent.

The angular structure of the outflow further illustrates the distinction between the relativistic jet and the baryon-loaded disk/propeller wind. The upper-middle and upper-right panels of Fig.~\ref{fig:sigma_jet_power_theta} show the time-averaged angular profiles of the isotropic-equivalent ``jet'' luminosity, $4\pi\,dL_{{\rm EM},\sigma>1}/d\Omega$, and magnetization $\sigma$, where $d\Omega=2\pi\sin\theta\,d\theta$. The isolated magnetar provides a useful reference: apart from the narrow low-$\sigma$ equatorial current sheet, its
outflow approximately follows the force-free aligned-rotator scaling
$dL_{\rm jet}/d\Omega\propto\sin^2\theta$ and remains highly magnetized, with $\sigma\simeq 20$, over most polar angles. Once an accretion disk is present, the outflow separates into the two components sketched in Fig.~\ref{fig:cartoon}: a magnetically dominated polar jet with $\sigma>1$ and a lower-$\sigma$, more equatorially concentrated disk/propeller wind. 

As $B_\star$ increases, the angular jet-power profile becomes both
brighter and broader, with its peak shifting to larger polar angle.
Consistent with this trend, Table~\ref{tab:LumJet} shows that the mean
jet luminosity rises from
$L_{\rm jet}\simeq5.8\times10^{50}\,{\rm erg\,s^{-1}}$ in
\texttt{NS-B1e15} to
$\simeq1.2\times10^{52}\,{\rm erg\,s^{-1}}$ in
\texttt{NS-B1e16}. The weaker-field accreting models therefore produce narrower polar
jets, possibly because of confinement by the inertia of the
baryon-loaded disk wind at lower latitudes. Additional collimation may
occur outside our simulation domain as the jet interacts with
surrounding merger or supernova ejecta
\citep[e.g.,][]{Bromberg+14,Gottlieb&Nakar22}.

The distribution of outflow power with magnetization is shown in the
lower-left panel of Fig.~\ref{fig:sigma_jet_power_theta}. In the
accreting NS models, most of the EM luminosity is carried by material
of moderate magnetization, $\sigma\lesssim 5$, but a non-negligible
high-$\sigma$ tail extends to $\sigma\sim10$--$20$. This
high-magnetization component becomes most prominent in the propeller
model \texttt{NS-B1e16}, reflecting the larger fraction of the outflow
power carried by the relativistic polar jet. The absolute normalization of $\sigma$ is floor-dependent and should not
be interpreted as a direct prediction of the physical jet magnetization.
In reality, the proto-magnetar wind mass loading is set by neutrino
heating from the stellar atmosphere and accretion flow, and may evolve
with the neutrino luminosity (e.g., as the magnetar cools); the simulated $\sigma$ distribution is
therefore most useful for comparing the relative degree of disk
entrainment and mixing across models.

The relative importance of the electromagnetic and matter components
also changes systematically with $B_\star$. Table~\ref{tab:LumJet}
gives the electromagnetic luminosity in $\sigma>1$ material, the total
electromagnetic luminosity, and the kinetic luminosity, all measured at
$r=1000\,r_{\rm g}$. In the crushed and channeled-accretion models,
most of the large-radius outflow power is carried by the baryon-loaded
matter component. For example, in \texttt{NS-B1e15},
$L_{\rm kin}\simeq10^{51}\,{\rm erg\,s^{-1}}$ exceeds
$L_{\rm EM,tot}\simeq9.4\times10^{49}\,{\rm erg\,s^{-1}}$ by about an
order of magnitude. As the stellar field is increased, the
electromagnetic component grows more rapidly than the kinetic
component. By \texttt{NS-B1e16}, the two are comparable, with
$L_{\rm EM,tot}\simeq6.5\times10^{51}\,{\rm erg\,s^{-1}}$ and
$L_{\rm kin}\simeq5.2\times10^{51}\,{\rm erg\,s^{-1}}$. Thus,
increasing $B_\star$ not only strengthens the relativistic polar jet,
but also changes the global outflow from a kinetic-energy-dominated
disk wind to a propeller/jet system with comparable electromagnetic and kinetic powers.

Finally, the lower-middle panel of Fig.~\ref{fig:sigma_jet_power_theta} shows how the kinetic power is distributed with asymptotic radial four-velocity, $(\gamma v)_{\infty}$. In the BH model, most of the outflow power is carried by sub-relativistic material with $(\gamma v)_{\infty}\sim 0.1$, consistent with previous simulations of neutrino-cooled BH accretion disks \citep[e.g.,][]{Siegel&Metzger18}. By contrast, the accreting NS models place a substantial fraction of their outflow energy into trans-relativistic ejecta with $(\gamma v)_{\infty}\sim 1$. In the propeller model \texttt{NS-B1e16}, the distribution extends beyond $(\gamma v)_\infty\sim1$ and reaches
values of order $10$, showing that rapidly rotating,
strongly magnetized NSs can simultaneously drive baryon-rich winds and
powerful relativistic jets.

\subsection{Torques on the Neutron Star}

The same disk--magnetosphere interaction that enhances the jet power
also determines whether the NS spins up or spins down.  
Accreting matter tends to add angular momentum to the star whereas the
magnetized polar jet extracts it. The star--disk magnetic connection can add or remove
angular momentum depending on whether the magnetically connected disk material rotates faster or
slower than the star, which roughly corresponds to connection inside
or outside the corotation radius. The net torque therefore depends on the competition
between hydrodynamic accretion through the surface, electromagnetic
stresses near the magnetosphere, and the Poynting-flux torque carried
by the relativistic jet.

We separate the torque into contributions associated with matter
crossing the stellar surface, magnetic stresses exerted locally by the
closed magnetosphere/accretion columns, and the angular momentum carried
away by the open-field relativistic jet.  This separation is useful
because these components are measured most cleanly on different
surfaces.  The jet torque is associated with the quasi-steady
Poynting-flux outflow on open polar magnetic field lines, and is
therefore measured through a spherical surface at the light-cylinder
radius.  By contrast, the accretion and closed-field magnetic torques
are applied directly to the star through the inner magnetosphere and
are measured just outside the neutron-star surface, where the accretion
columns and closed star--disk field lines intersect the inner boundary.
We therefore define the total stellar torque as the sum of three
bookkeeping components: the matter torque through the surface, the
near-surface electromagnetic torque after removal of the open-field jet
contribution, and the jet torque carried away along open field lines.

Operationally, we compute the angular-momentum fluxes as follows.  The
hydrodynamic torque is
\begin{equation}
\tau_{\rm hydro} =
-\int \rho h u^r u_\phi \sqrt{-g}\,d\theta d\phi,
\label{eq:hydro_torque}
\end{equation}
while the electromagnetic torque is
\begin{equation}
\tau_{\rm EM} =
-\int \left(b^2 u^r u_\phi - b^r b_\phi\right)
\sqrt{-g}\,d\theta d\phi .
\label{eq:EM_torque}
\end{equation}
For the jet contribution, we evaluate Eq.~\eqref{eq:EM_torque} at
$R_{\rm LC}$ and restrict to the magnetically dominated open-field
outflow.  This gives the angular momentum loss associated with the
large-scale Poynting flux, which is comparatively steady and less
sensitive to the highly time-dependent dynamics of the inner
magnetosphere.  For the surface contribution, we evaluate the same stress tensor just outside the inner boundary and subtract the contemporaneous jet torque, thereby isolating the remaining ``EM non-jet'' torque from closed field lines, accretion columns, and the disk--magnetosphere connection.

When the magnetosphere is pushed close to the NS surface in the
accretion regime, dynamic plasmoid eruptions occur
(Fig.~\ref{fig:plasmoid}), creating transient regions with large
positive radial velocity near the inner boundary.  These regions are
part of the time-dependent magnetospheric outflow rather than material
being accreted by the star.  To avoid counting them as hydrodynamic
accretion torque, we include in $\tau_{\rm hydro}$ only grid cells with
negative radial velocity.  When calculating the near-surface EM torque,
we retain both terms in Eq.~\eqref{eq:EM_torque} in zones with negative
radial velocity.  In zones with positive radial velocity, we retain
only the Maxwell stress term, $-b^r b_\phi$, since this represents the
magnetic stress exerted on the star even when the plasma itself is
moving outward.

\begin{figure*}
\centering
\safeincludegraphics[width=\textwidth]{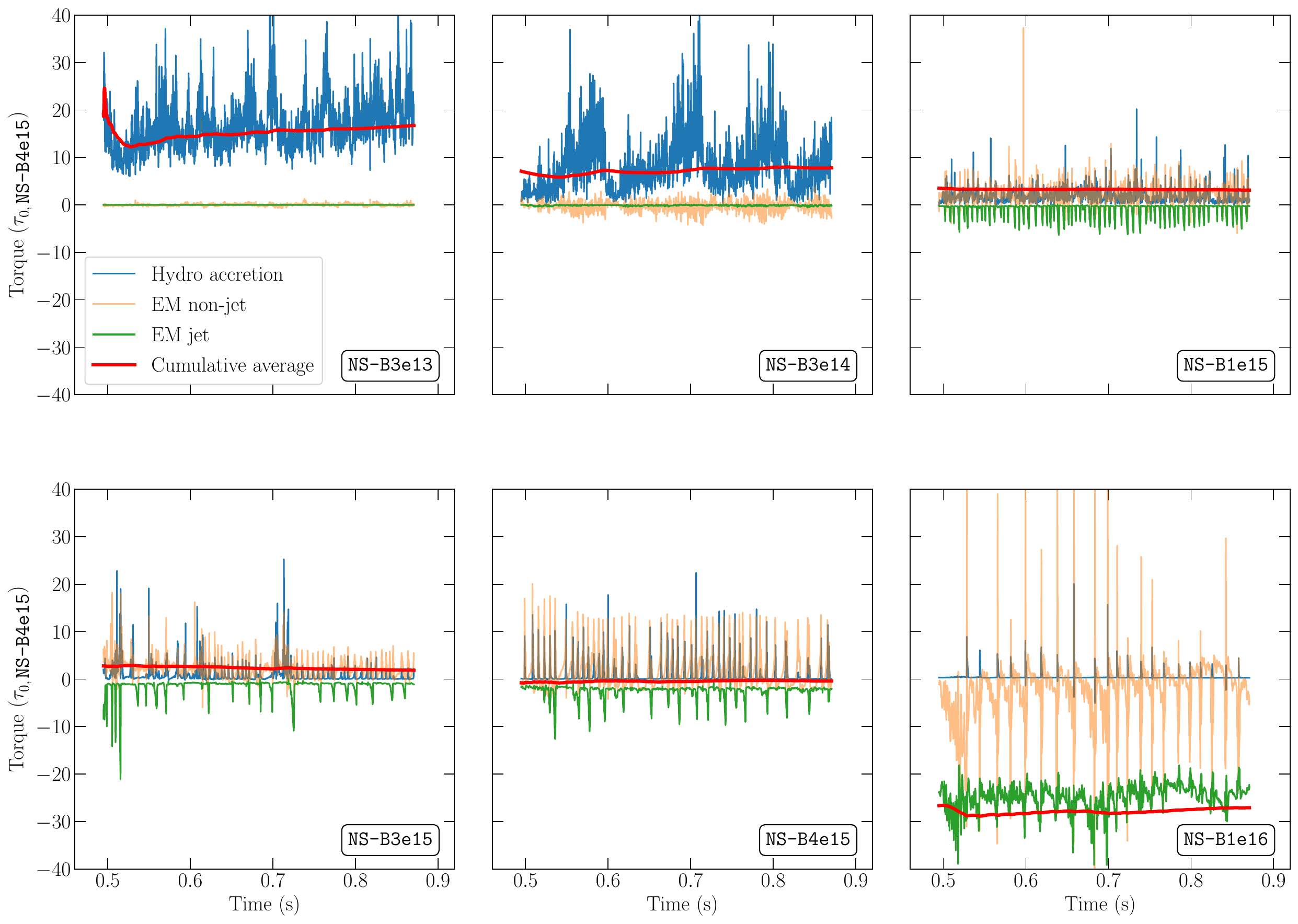}
\caption{Contributions to the neutron-star torque as a function of
time, normalized to the magnitude of the isolated-dipole torque for
$B_{\star}=4\times10^{15}\,{\rm G}$. Blue curves show the
hydrodynamic torque from matter crossing the stellar surface, orange
curves show the near-surface electromagnetic torque excluding the jet,
and green curves show the spin-down torque carried by the
magnetically dominated jet. Each panel is labeled by the neutron-star
model. Thick red curves show the cumulative average of the total
torque. Positive values indicate spin-up and negative values
spin-down.}
\label{fig:torque_plot}
\end{figure*}

Fig.~\ref{fig:torque_plot} shows the time evolution of these torque
contributions, normalized to the magnitude of the dipole spin-down
torque of an otherwise equivalent isolated NS at
$B_\star=4\times10^{15}\,{\rm G}$,
$\tau_0=\dot{E}_0/\Omega_\star$. Unlike
Table~\ref{tab:LumJet}, which normalizes each model to its own
isolated-dipole torque, Fig.~\ref{fig:torque_plot} uses this single
reference value so that the absolute torque amplitudes can be compared
across the model sequence. Negative torques indicate spin-down, while
positive torques indicate spin-up. Blue curves denote the
hydrodynamic torque from matter accreted through the inner boundary,
green curves denote the jet torque measured from the open-field
Poynting flux at $R_{\rm LC}$, and orange curves denote the remaining
near-surface electromagnetic torque after subtracting the jet
contribution, i.e., the ``EM non-jet'' torque from closed field lines,
accretion columns, and the local disk--magnetosphere interaction. The
thick red curve in each panel shows the cumulative time average of the
total torque.

The net torque changes systematically across the model sequence. In the
weakest-field models \texttt{NS-B3e13} and \texttt{NS-B3e14}, the
stellar field is dynamically unimportant and hydrodynamic accretion
strongly spins up the star. Model \texttt{NS-B1e15} also spins up, with
a mean torque approximately twenty times the magnitude of the
corresponding isolated dipole spin-down torque
(Table~\ref{tab:LumJet}). As the field is increased, the negative jet
and magnetospheric torques grow while the hydrodynamic spin-up torque
declines. In \texttt{NS-B3e15}, the net torque remains positive but is
substantially reduced. By \texttt{NS-B4e15}, the positive accretion
torque and negative electromagnetic torques nearly cancel, placing the
system close to spin equilibrium; its time-averaged torque is slightly
negative but consistent with zero relative to its large temporal
fluctuations. This near cancellation is consistent with
$\langle R_{\rm m}\rangle\simeq R_{\rm c}$: for
\texttt{NS-B4e15}, the two estimates give
$\langle R_{\rm m}\rangle\simeq6.0$--$6.5\,r_{\rm g}$, close to
$R_{\rm c}=7.3\,r_{\rm g}$ (Table~\ref{tab:models}). In
\texttt{NS-B1e16}, the magnetosphere lies outside corotation, little
matter crosses the centrifugal barrier, and the star spins down
electromagnetically. The spin-down torque remains enhanced by roughly a factor of 2 relative to an isolated magnetar because the disk opens
magnetic flux that would otherwise remain closed. The resulting
spin-evolution timescale ranges from several tens of seconds in the
accreting spin-up models, to hundreds of seconds near torque balance,
and to only about five seconds in the propeller model
(Table~\ref{tab:LumJet}).

\section{Application to GRB Central Engines}
\label{sec:applications}

\subsection{Collapsars}

As a concrete application of our findings, consider the collapse of a rotating helium core of mass $\sim 3-10\,M_{\odot}$, the canonical stripped-envelope progenitors of most long-duration GRBs (\citealt{Woosley&Bloom06}).  Rather than immediately forming a black hole, the iron core of the collapsing star first produces a massive, rapidly rotating proto-neutron star (e.g., \citealt{Dessart+08}).  We assume that the NS acquires, likely through an MHD dynamo powered
by differential rotation \citep{Kiuchi+24}, a large-scale dipole field
$B_\star\sim{\rm few}\times10^{15}\,{\rm G}$, comparable to the
near-corotation model \texttt{NS-B4e15} studied below and somewhat
stronger than the inferred dipole fields of most Galactic magnetars.

The outer layers of the stellar core, with greater angular momentum, form an accretion disk around the magnetar, which feeds it at a characteristic outer rate $\dot{M}_{\rm d} \gtrsim 0.1\,M_{\odot}\,{\rm s^{-1}}$ similar to those sampled by our simulations, for a duration $t_{\rm acc} \sim 10-30\,{\rm s}$, as set by the infall time of the stellar core (e.g., \citealt{Kumar+08}; though, as we discuss below, not all of this inflowing mass is ultimately accreted by the magnetar).  This timescale being comparable to the observed durations of long GRBs \citep{vonKienlin20} is one of the chief motivations for the collapsar model \citep{MacFadyen&Woosley99}.  

We focus attention on the near-corotation model
\texttt{NS-B4e15}, because it lies closest to spin equilibrium at the
adopted accretion rate, with
$\langle R_{\rm m}\rangle\simeq R_{\rm c}$. In this model, the positive
accretion torque and negative electromagnetic torques nearly cancel,
giving a time-averaged torque that is slightly negative but consistent
with zero relative to its large temporal fluctuations
(Table~\ref{tab:LumJet}). More generally, if the accretion rate and stellar magnetic field evolve
slowly compared with the torque-adjustment time, an accreting magnetar
will tend toward an approximate equilibrium spin for which
$R_{\rm m}\sim R_{\rm c}$. A star initially rotating
more slowly than this equilibrium value will tend to spin up through
accretion, whereas a more rapidly rotating star will tend to spin down
through its magnetosphere and relativistic jet.

Although the disk supplies material at an outer rate
$\dot{M}_{\rm d}\gtrsim0.1\,M_\odot\,{\rm s^{-1}}$, only a fraction of
this material reaches the stellar surface. For
\texttt{NS-B4e15}, the time-averaged stellar accretion rate is
$\dot{M}_{\star}\simeq8\times10^{-3}\,M_\odot\,{\rm s^{-1}}$
(Fig.~\ref{fig:radial_profiles}; Table~\ref{tab:models}). If this time-averaged rate and magnetospheric state were maintained for
$t_{\rm acc}\sim10$--$30\,{\rm s}$, the neutron star would accrete
$\Delta M_{\star}\sim\dot{M}_{\star}t_{\rm acc}
\simeq0.08$--$0.24\,M_\odot$.  Thus, provided the initial proto-neutron star lies more than a few
tenths of a solar mass below its time-dependent maximum supported mass, it could survive throughout the active accretion phase without prompt collapse to a black hole.

For the same model, the mean relativistic jet power is
$L_{\rm jet}\simeq1.1\times10^{51}\,{\rm erg\,s^{-1}}$
(Table~\ref{tab:LumJet}), broadly within the range required of
long-GRB central engines. For a prompt radiative efficiency
$\epsilon_\gamma\sim0.1$, this corresponds to a beaming-corrected
gamma-ray luminosity
$L_\gamma=\epsilon_\gamma L_{\rm jet}\sim10^{50}\,{\rm erg\,s^{-1}}$,
comparable to observationally inferred beaming-corrected GRB
luminosities \citep{Frail+01,Yi+17}.

Previous studies of isolated proto-magnetar spin-down show that the jet
magnetization, as set by neutrino-driven baryon loading near the stellar
surface, can rise rapidly as the magnetar cools over a timescale of tens
of seconds \citep{Metzger+11a}. Such strong secular evolution may be in
tension with time-resolved GRB observations, which generally show more
moderate evolution of the high-energy prompt-emission properties over
the duration of the burst
\citep{Metzger+11a,Beniamini+17,Uhm2018}. In an accreting
proto-magnetar, however, the accretion flow supplies an additional,
comparatively steady source of neutrinos. For the near-corotation model
\texttt{NS-B4e15}, the volume-integrated electron-flavor neutrino
luminosity generated within the simulated accretion flow is of order
$L_{\nu_{\rm e}}+L_{\bar{\nu}_{\rm e}}\sim10^{51}\,
{\rm erg\,s^{-1}}$ (Fig.~\ref{fig:neutrino_lumtemp}). This accretion-powered component can become comparable to or exceed the declining
internal cooling luminosity of the proto-neutron star after the first
several seconds. Irradiation of the polar cap by the accretion flow could therefore
maintain a more nearly constant baryon loading and jet magnetization
than in the isolated-magnetar case \citep{Metzger+18b}, potentially
reducing the secular evolution of the jet's high-energy emission.

Over the same characteristic interval of tens of seconds, the
large-radius electromagnetic and bulk kinetic components carry a
combined power
$L_{\rm EM,tot}+L_{\rm kin}\simeq1.5\times10^{51}\,{\rm erg\,s^{-1}}$
for \texttt{NS-B4e15} (Table~\ref{tab:LumJet}). If sustained for
$t_{\rm acc}\sim10$--$30\,{\rm s}$, these components alone would
inject an energy of order $10^{52}\,{\rm erg}$ into the surrounding
star. This partitioning of the engine
power is qualitatively similar to that inferred for GRB-supernovae, in
which the kinetic energy of the accompanying broad-lined Type Ic
supernova, $\sim10^{52}\,{\rm erg}$, typically exceeds the
beaming-corrected GRB gamma-ray energy, $\sim10^{51}\,{\rm erg}$
\citep{Frail+01,Cano+17,Gottlieb+25b}.

\subsection{Mergers and AIC}

Closely related considerations apply to accreting neutron-star remnants formed in the AIC of rapidly spinning white dwarfs and in neutron-star mergers, which are possible central engines of short-duration GRBs \citep[e.g.,][]{Usov92,Dessart+06,Metzger+08a,Bucciantini+12,combi_aic_2025,Cheong+25,Gottlieb+25b}.  In these environments, the characteristic disk masses $\sim 10^{-2}-0.1\,M_{\odot}$ and hyper-accretion rates $\dot{M}_{\rm d} \sim 0.1M_{\odot}$ s$^{-1}$ overlap those of our simulations and those achieved in collapsars.  One key difference is that a neutron-star-merger remnant is generally
formed much closer to its maximum supported mass than an AIC remnant.
Its lifetime may therefore be controlled not only by accretion from the
surrounding disk, but also by internal redistribution of thermal energy
and angular momentum \citep{Margalit+22}. Nevertheless, while the remnant survives as a rapidly rotating,
strongly magnetized NS, the local disk--magnetosphere interaction
should still be governed primarily by the ordering of
$R_\star$, $R_{\rm m}$, $R_{\rm c}$, and $R_{\rm LC}$. Thus, even though the global environment differs from collapsars, the
local disk--magnetosphere interaction can redistribute the engine power
between a relativistic jet and a baryon-rich outflow in much the same
way.

For a magnetar to power a short-duration GRB, its polar outflow must
become sufficiently relativistic within the first few seconds after
formation, comparable to the observed prompt-emission duration
$t_\gamma\lesssim2\,{\rm s}$.  Recent GRMHD simulations of young millisecond magnetars including neutrino-transport suggest that comparatively baryon-clean polar funnels with magnetizations as large as $\sigma \sim 10$ can be achieved even in the presence of strong neutrino-driven or magnetically driven mass loss present within the first second after formation \citep{Combi&Siegel23,Desai+26}.  If such conditions are realized, the polar component could power a short-duration GRB, while the accompanying disk/propeller outflow would add mass and energy to the kilonova ejecta.  For otherwise similar disk masses, the resulting ejecta mass and
kinetic energy could exceed those expected following prompt black-hole
formation, where the outflow is powered only by the black-hole--torus
system \citep{Siegel&Metzger17}.  This additional outflow component could therefore modify both the luminosity and color evolution of the kilonova, as well as the amount of neutron-rich material available for $r$-process nucleosynthesis (see Paper II for further discussion).  The coexistence of a relativistic jet with a more broadly distributed,
baryon-rich outflow in the near-equilibrium models is qualitatively
compatible with events in which substantial energy is inferred in both
the short-GRB jet and the kilonova ejecta
\citep{rastinejad_kn_grb_2025,Gottlieb+25b}.

\section{Summary and Conclusions}
\label{sec:conclusions}

\begin{table*}
\centering
\caption{Summary of accretion regimes for proto-magnetar disk interaction.}
\label{tab:accretion_regimes}
\begin{tabular}{llll}
\hline
Regime & Ordering of radii & Properties & Models \\
\hline
Crushed magnetosphere &
$R_{\rm m} \lesssim R_{\star} < R_{\rm c}$ &
Magnetosphere compressed to the stellar surface; 
& \texttt{NS-B3e13} \\
& & Inner flow resembles that of a weakly magnetized accretor; 
& \texttt{NS-B3e14} \\
& & Compressed stronger-field cases can exhibit plasmoid-driven jet
variability.
& \texttt{NS-B1e15} \\
\hline
Channeled accretion &
$R_{\star} < R_{\rm m} \lesssim R_{\rm c}$ & 
Accretion channeled along closed magnetic field lines; 
& \texttt{NS-B3e15} \\
& & System approaches spin equilibrium as
$R_{\rm m}\rightarrow R_{\rm c}$;  
& \texttt{NS-B4e15} \\
& & Enhanced baryon-rich disk outflows (Paper II). & \\
\hline
Propeller &
$R_{\star} < R_{\rm c} < R_{\rm m}$ & 
Centrifugal coupling suppresses direct accretion; 
& \texttt{NS-B1e16} \\
& & Baryon-rich propeller outflow; & \\
& & Rapid electromagnetic spin-down; & \\
& & Reduced jet variability. & \\
\hline
\end{tabular} 
\end{table*}

We have presented axisymmetric GRMHD simulations of hyperaccretion
onto rapidly rotating, magnetized neutron stars, including a physical
equation of state and charged-current weak interactions. Rather than
modeling a complete collapsar, neutron-star merger, or accretion-induced
collapse event, we isolate the local disk--magnetosphere interaction in
a controlled numerical experiment. Holding fixed the initial torus and
compact-object properties, we vary the neutron-star dipole field strength from
$3\times10^{13}$ to $10^{16}\,{\rm G}$, following the progression from
a crushed magnetosphere through channeled accretion and approximate
spin equilibrium to propeller-driven spin-down, and compare this
sequence with an otherwise similar accreting black hole.  This strategy allows us to separate the effects of the stellar surface
and magnetosphere from those of the accretion flow itself.

Our calculations focus on a SANE-like disk whose magnetic field is weak
compared with the stellar dipole. The chosen parameters place the inner
disk above the neutronization threshold while resolving the stellar,
magnetospheric, corotation, and light-cylinder radii. Although the
specific field strengths and accretion rate are tailored to this setup,
the magnetospheric regimes are approximately scalable: at fixed
$R_{\rm m}/r_{\rm g}$ and stellar parameters,
$B_\star\propto\dot M^{1/2}$, or equivalently
$\dot M\propto B_\star^2$.  In contrast to black-hole accretion in the SANE regime, the accreting
proto-magnetars generically launch both strong baryon-rich outflows with
$\gamma\beta\sim1$ and ultrarelativistic, magnetically dominated jets.

Increasing the stellar dipole strength produces a continuous transition
from a crushed magnetosphere, through magnetically channeled accretion
and near-corotation torque balance, to a centrifugal propeller. Even
before a distinct magnetospheric cavity forms, a sufficiently strong
compressed field can modify the near-surface inflow and enhance the
outflows. As $R_{\rm m}$ approaches $R_{\rm c}$, direct accretion becomes
increasingly intermittent; once $R_{\rm m}>R_{\rm c}$, it is
suppressed and a larger fraction of the supplied mass is redirected
into disk and magnetospheric outflows.

Accretion opens additional stellar magnetic flux inside the light
cylinder and increases the relativistic jet power above the isolated
dipole value \citep{Parfrey+16,Metzger+18b}. For the
magnetar-strength models most relevant to GRB engines, the mean jet
power is enhanced by factors of approximately $2$--$11$. Even in the
propeller regime, where direct accretion is suppressed, the
jet remains about twice as powerful as that of an isolated dipole with
the same field and spin. Isolated-dipole estimates can therefore
substantially underpredict the power of an accreting proto-magnetar.

The disk--magnetosphere interaction also makes the jet strongly
time-variable. Channeled-accretion and near-corotation models undergo
reconnection-driven plasmoid eruptions, whereas the propeller model
produces a steadier jet. As the fallback rate evolves in a collapsar or
the disk drains in a merger or AIC event, transitions among crushed,
channeled, near-equilibrium, and propeller states may therefore imprint
additional variability on the central engine
\citep{Piro&Ott11,Gompertz+14,Metzger+18b}.

The stellar torque changes qualitatively across the same sequence:
weak-field models spin up, near-corotation models approach torque
balance, and propeller models spin down rapidly. Thus, the accretion
regime controls not only the jet power and variability, but also the
secular spin evolution of the neutron star.

The magnetosphere also regulates how efficiently the neutron star gains
mass. In the propeller regime, most of the gas supplied at large radii
fails to reach the surface, while even channeled and near-corotation
models redirect a substantial fraction of the inflow into outflows.
The lifetime of an accreting proto-magnetar may therefore depend not
only on the external fallback supply, but also on its instantaneous
magnetospheric regime. Inefficient accretion can delay collapse to a
black hole while the system continues to power relativistic and
baryon-rich outflows.  The composition and nucleosynthetic implications of the baryon-rich
disk and propeller outflows are explored in Paper~II.

Applied to collapsars, our results suggest that a millisecond
proto-magnetar with a dipole field of a few
$\times10^{15}\,{\rm G}$ can reside near spin equilibrium while
powering a long-GRB jet. Because only a fraction of the externally supplied mass reaches the
surface, such a neutron star may avoid prompt collapse for tens of
seconds while its relativistic jet and baryon-rich outflows inject an
energy comparable to that inferred for GRB-supernovae. Accretion-powered neutrino irradiation may also
moderate the secular evolution of the jet baryon loading relative to an
isolated proto-magnetar.  These results support accreting proto-magnetars as viable central
engines for long GRBs and energetic explosions.

A major assumption is that the newly formed neutron star amplifies and
organizes a large-scale field $\gtrsim10^{15}\,{\rm G}$ before
unimpeded early accretion drives it to collapse. A comparable assumption---the presence of a strong, ordered magnetic
flux threading the central object---is also required in black-hole GRB
models powered by the Blandford--Znajek mechanism
\citep{Gottlieb+22}, although the relevant field-amplification
timescale may be less restrictive in the black-hole case.

Our results also connect to simulations of isolated, neutrino-heated
proto-magnetar winds. During the first seconds after birth, rapid
rotation can produce a baryon-rich equatorial wind while leaving the
polar field lines comparatively clean
\citep{Combi&Siegel23,Desai+26}. The disk-fed systems studied here
provide a later-time realization of the same basic geometry: a
high-magnetization polar jet confined by a more baryon-rich,
lower-latitude outflow. The transition from an early
neutrino-wind-dominated phase to a later disk-fed phase may therefore
maintain rather than disrupt the jet and its collimating environment.

\subsection{Limitations and Future Work}

Several limitations of this work motivate future study.  First, we have focused on disks with relatively weak magnetic fields.  A stronger disk field, such as might be produced by a dynamo associated with the MRI \citep{JacqueminIde+24,Chan+26}, could change the amount of magnetic flux delivered to the central object, the degree of jet collimation, and the reconnection geometry at the disk--magnetosphere interface.  In neutron-star accretion, the relative polarity between the disk field and the stellar dipole introduces an additional degree of freedom.  \citet{Parfrey&Tchekhovskoy24} found in 3D GRMHD simulations that the jet power and inner-disk structure can depend on whether the stellar and disk fields are parallel or anti-parallel, especially when the disk penetrates well inside the corotation radius, whereas the polarity dependence weakens in the propeller regime.  Future simulations should explore how such polarity effects operate in neutrino-cooled proto-magnetar disks.

Second, our simulations are axisymmetric.  Three-dimensional effects can alter both the accretion flow and the reconnection dynamics.  In particular, \citet{Zhu+24} found that plasmoid eruptions can be overestimated in two dimensions because interchange instabilities in three dimensions provide an additional source of effective resistivity, allowing the disk and stellar fields to reconnect in a more quasi-steady manner.  Axisymmetry may exaggerate the intermittency of disk--magnetosphere
reconnection: in 3D, non-axisymmetric fingers can penetrate the
magnetosphere and provide more continuous accretion or mass leakage
through the stellar field.  Moreover, propagation through the stellar
envelope or merger/AIC ejecta can substantially reshape any variability
imprinted by the engine before radiation is produced at larger radii (e.g., \citealt{Gottlieb+22}). At the same time, three-dimensional simulations are required to capture the full development of the MRI, the possibility of a disk dynamo, non-axisymmetric mass loading of the magnetosphere, and the stability of the jet--wind interface.  Extending the present calculations to 3D, with neutrino microphysics and a physical equation of state, will therefore be essential for determining whether the episodic jet variability found here persists in more realistic settings.
It is also necessary to break axisymmetry to investigate the effect of a nonzero angle between the star's spin and magnetic axes \citep[e.g.][]{Romanova+03}, which can reduce the relative jet-power enhancement due to accretion \citep{Murguia-Berthier2024,Das2024}.

Third, our treatment of the neutron-star surface is idealized. We impose an artificial inner boundary at the stellar surface and therefore do not resolve the proto-neutron-star atmosphere, where the baryon loading of open magnetic field lines is physically set. In reality, the mass flux emerging from the surface is determined by the balance between neutrino heating from the cooling proto-neutron star and neutrino cooling in the subsonic portion of the wind, with additional latitude-dependent modifications from centrifugal support along rotating field lines \citep{Metzger+07,Desai+26}. As a result, the angular structure of the jet magnetization in our simulations, shown in Fig.~\ref{fig:sigma_jet_power_theta}, reflects only baryon loading produced by interaction and mixing between the polar magnetosphere and the accretion flow. In a physical proto-magnetar, the same $\sigma(\theta)$ structure would also depend on the intrinsic, latitude-dependent mass loading from the neutron-star atmosphere. Resolving this atmosphere, together with neutrino heating and cooling at the stellar surface, will be necessary to predict the absolute magnetization of the jet and its angular dependence throughout the magnetar's cooling evolution.

\section*{Acknowledgements}
T.~P.~ and B.~D.~M. gratefully acknowledge support from the National Science Foundation (grant AST-2406637), NASA (grants 80NSSC24K0408,80NSSC26K0299,80NSSC22K0807), the Gordon and Betty Moore Foundation (grant 13920; PI Ruben Gonzalez), and the Simons Foundation (grant 727700).  The Flatiron Institute is supported by the Simons Foundation. KP acknowledges support from the Laboratory Directed Research and Development Program at Princeton Plasma Physics Laboratory, a national laboratory operated by Princeton University for the U.S.\ Department of Energy under Prime Contract No.\ DE-AC02-09CH11466 and from NASA grant 80NSSC21K1746. We thank Rodrigo Fernandez and Yixian Chen for helpful discussions on microphysics related aspects. 

\appendix





\section{Checking the optically thin assumption}
\label{sec:optically_thin}

\begin{figure*}
\centering
\includegraphics[width=\textwidth]{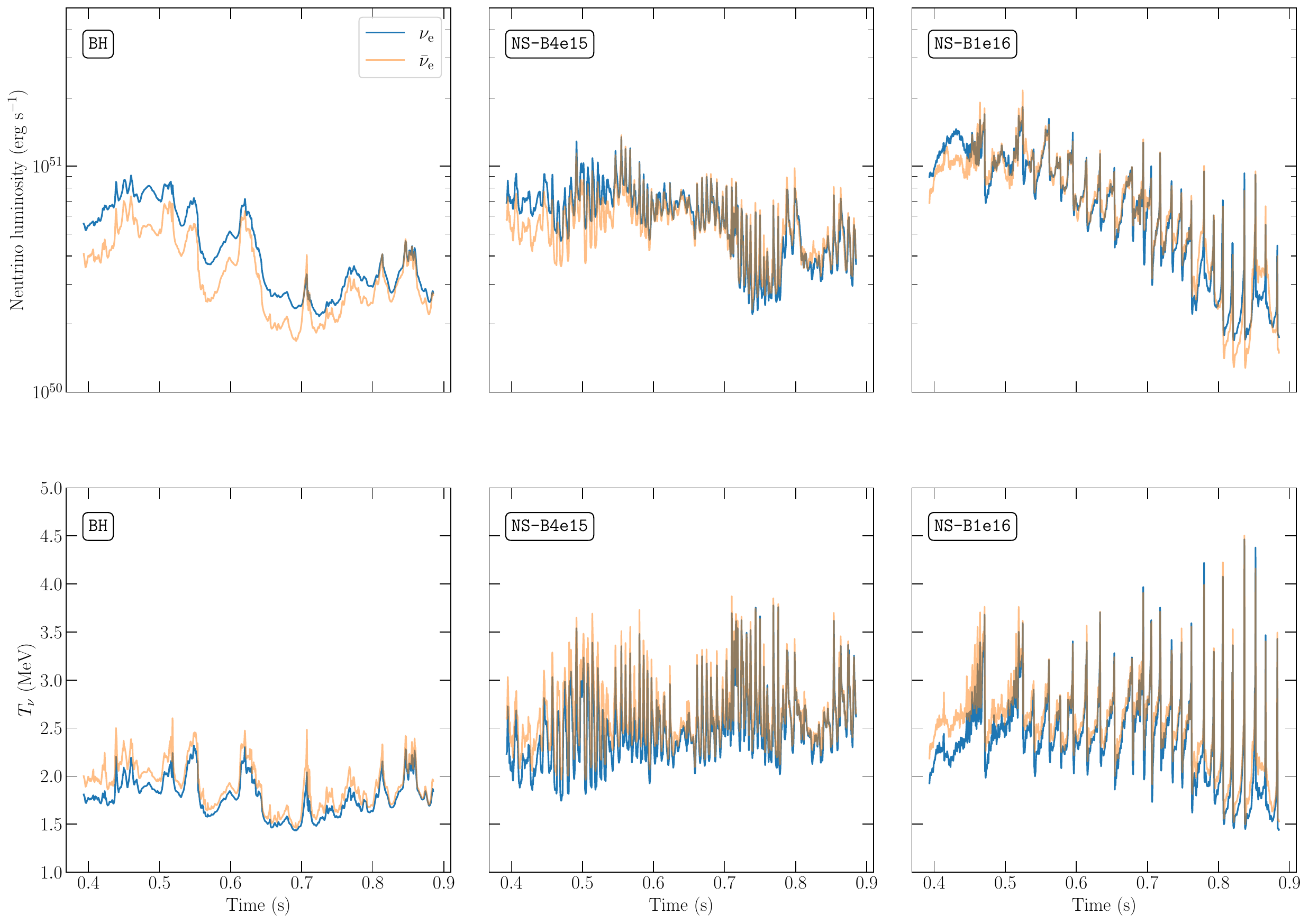}
\caption{Electron-neutrino and electron-antineutrino luminosities for
the BH and NS models (top panels), and the corresponding
luminosity-weighted average emission temperatures (bottom panels;
Eq.~\eqref{eq:Tnu}).}
\label{fig:neutrino_lumtemp}
\end{figure*}

\begin{figure*}
\centering
\includegraphics[width=\textwidth]{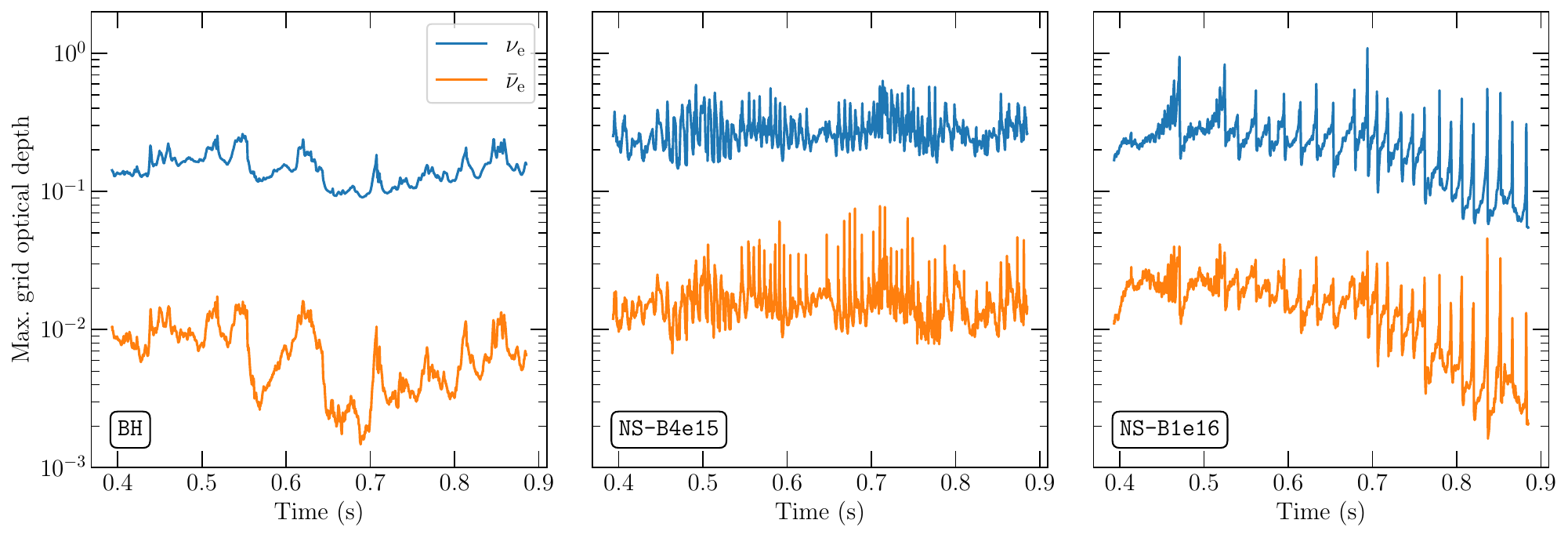}
\caption{Maximum local optical depth to infinity across the grid as a
function of time for electron neutrinos and electron antineutrinos in
the BH and NS models.}
\label{fig:optical_depth}
\end{figure*}

Our simulations neglect neutrino transport and assume optically thin
neutrino cooling (Sec.~\ref{sec:weak}). Here we verify that this
approximation is adequate for the models presented in this paper.
Fig.~\ref{fig:neutrino_lumtemp} shows that the accretion flows
produce electron-flavor neutrino luminosities
$L_{\nu_{\rm e}}+L_{\bar{\nu}_{\rm e}}\sim 10^{51}\,{\rm erg\,s^{-1}}$,
with luminosity-weighted emission temperatures of a few MeV. Despite
these high luminosities, Fig.~\ref{fig:optical_depth} shows that the
maximum optical depth to infinity remains below unity for nearly all
times and locations. The only brief excursions to
$\tau_{\nu_e,{\rm abs}}\sim1$ occur in a small number of cells close to
the NS surface, where localized density and temperature spikes form near the
disk--magnetosphere interface in the propeller regime.

\begin{figure*}
\centering
\includegraphics[width=\linewidth]{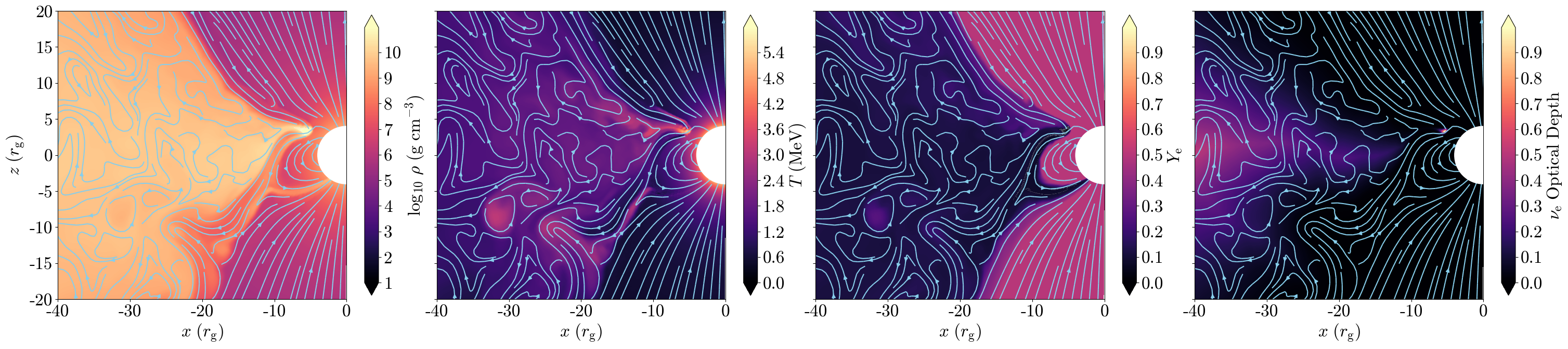}
\caption{Two-dimensional maps of density, temperature, electron
fraction, and local optical depth near the NS surface at one of the
brief instances when the maximum optical depth on the grid spikes to
$\sim 1$ in the propeller regime (right panel of
Fig.~\ref{fig:optical_depth}).}
\label{fig:optical_depth_2d}
\end{figure*}

We estimate the minimum optical depth for neutrinos to escape from
each location in the simulation domain using a local path-minimization
procedure. The optical depth from a given cell to infinity is written
as the optical depth from infinity to a neighboring cell plus the optical depth
between the two cells \citep{Neilsen2014,Siegel&Metzger18}. We first
set the optical depth to zero in all cells with density below
$10^{-6}\rho_{\rm max}$, where $\rho_{\rm max}$ is the maximum grid
density, and initialize the remaining cells with infinite optical
depth. We then sweep across the grid and update the optical depth in
each cell according to
\begin{equation}
\tau_{\rm cell} =
\min\left[\kappa\,ds + \tau_{\rm neighbor}\right],
\label{eq:taucell}
\end{equation}
where the minimization is performed over neighboring cells whose
optical depths have already been updated, $ds$ is the distance between
cell centers computed using the metric averaged between them, and
$\kappa$ is the opacity between the two cells. To account for different
escape directions, we perform four sweeps corresponding to the
different combinations of radially inward/outward and north/south
motion. After these sweeps, each cell contains an estimate of the
minimum optical depth from that cell to infinity.

For this diagnostic, we include charged-current absorption of electron
neutrinos and antineutrinos on free nucleons, corresponding to the inverse reactions
of Eq.~\eqref{eq:URCA}. We neglect neutral-current
scattering, so the resulting optical depths should be interpreted as
absorption optical depths rather than complete transport optical depths. The charged-current opacities are
\begin{eqnarray}
\label{eq:optical_depth}
\kappa_{\bar{\nu}_{\rm e}} & = &
\left(\frac{Y_{\rm e}-X_{\alpha}/2}{m_{\rm n}}\right)
\sigma \rho; \,\,\,\,\, \kappa_{\nu_{\rm e}} =
\left(\frac{1-Y_{\rm e}-X_{\alpha}/2}{m_{\rm n}}\right)
\sigma \rho ,
\end{eqnarray}
where $X_\alpha$ is the mass fraction of $\alpha$ particles,
$m_{\rm n}$ is the nucleon mass, and
$\sigma=9\times10^{-44}\epsilon_{\nu,\rm MeV}^{2}\,{\rm cm^2}$ is the
neutrino absorption cross section. The quantities
$\rho(Y_{\rm e}-X_\alpha/2)$ and
$\rho(1-Y_{\rm e}-X_\alpha/2)$ are averaged between neighboring cells.
 We estimate a representative neutrino energy as
$\epsilon_{\nu,{\rm MeV}}=5.1T_{\nu,{\rm MeV}}$, appropriate for
optically thin pair-capture emission
\citep{Chen&Beloborodov07}. Because the opacity scales as
$\epsilon_\nu^2$, this grey estimate does not capture the full neutrino
energy distribution and should be regarded as an order-of-magnitude
optical-depth diagnostic. The luminosity-weighted emission
temperature is
\begin{equation}
T_{\nu} =
\frac{\int T Q_{\nu}\gamma \sqrt{-g}\, d^3x}
{\int Q_{\nu}\gamma \sqrt{-g}\, d^3x},
\label{eq:Tnu}
\end{equation}
where $T$ is the local gas temperature, $Q_\nu$ is the neutrino cooling
rate, and $\gamma$ is the Lorentz factor. The neutrino luminosity shown
in Fig.~\ref{fig:neutrino_lumtemp} is computed as
\begin{equation}
L_{\nu}= \int \alpha Q_{\nu}\gamma \sqrt{-g}\, d^3x,
\end{equation}
where $\alpha$ is the lapse function. 

When computing $T_\nu$, we exclude radii $r<6\,r_{\rm g}$ in the
channeled-accretion models. A small number of localized,
surface-adjacent cells reach very high temperatures when matter impacts
the inner boundary along magnetic field lines; because the stellar
surface and cooling layer are not resolved physically, including these
cells would allow an inner-boundary artifact to dominate the inferred
characteristic neutrino energy.

The resulting optical depths are shown in
Fig.~\ref{fig:optical_depth}. In all models, the flow remains optically
thin to both $\nu_{\rm e}$ and $\bar{\nu}_{\rm e}$ over essentially the
entire simulation. The largest values occur for $\nu_{\rm e}$ in the
propeller model with $B_\star=10^{16}\,{\rm G}$, where brief
spikes push the maximum optical depth close to unity. These spikes arise from localized density and temperature enhancements
when a small amount of matter penetrates the centrifugal barrier and
reaches the vicinity of the NS surface. Fig.~\ref{fig:optical_depth_2d} shows that the corresponding
$\tau_{\nu_{\rm e}}\sim 1$ region occupies only a small number of cells,
while the surrounding flow has much smaller optical depth.

We therefore conclude that the accretion flows in the present
simulations are optically thin to charged-current absorption over
essentially the entire domain. Although additional scattering opacity
could modestly increase the total transport optical depth in the small,
transient regions where $\tau_{\nu_e,{\rm abs}}\sim1$, these regions
occupy only a few cells and are unlikely to qualitatively affect the
global disk--magnetosphere interaction studied here.


\bibliography{refs}

\end{document}